\documentclass[pdflatex,sn-nature]{sn-jnl}

\usepackage{graphicx}%
\usepackage{multirow}%
\usepackage{amsmath,amssymb,amsfonts}%
\usepackage{amsthm}%
\usepackage{mathrsfs}%
\usepackage[title]{appendix}%
\usepackage{xcolor}%
\usepackage{textcomp}%
\usepackage{manyfoot}%
\usepackage{booktabs}%
\usepackage{algorithm}%
\usepackage{algorithmicx}%
\usepackage{algpseudocode}%
\usepackage{listings}%
\theoremstyle{thmstyleone}%
\theoremstyle{thmstyletwo}%

\theoremstyle{thmstylethree}%

\begin{document}

\title[Article Title]{Altermagnetic–doping interplay as a route to enhanced d-wave pairing in the Hubbard model}


\author*[1,2,3]{\fnm{Ji} \sur{Liu}}
\author*[4]{\fnm{Jianyu} \sur{Li}}
\author[1,3]{\fnm{Peng} \sur{Zhang}}
\author[2]{\fnm{Xiaosen} \sur{Yang}}\email{yangxs@ujs.edu.cn}
\author[1,3]{\fnm{Ho-Kin} \sur{Tang}}\email{denghaojian@hit.edu.cn}

\affil[1]{\orgdiv{School of Science}, \orgname{Harbin Institute of Technology}, \orgaddress{ \city{Shenzhen}, \postcode{518055},  \country{China}}}

\affil[2]{\orgdiv{Department of Physics}, \orgname{Jiangsu University}, \orgaddress{ \city{Zhenjiang}, \postcode{212013}, \country{China}}}


\affil[3]{\orgname{Shenzhen Key Laboratory of Advanced Functional Carbon Materials Research and Comprehensive Application}, \orgaddress{ \city{Shenzhen}, \postcode{518055},  \country{China}}}

\affil[4]{\orgdiv{Institute for Materials Research}, \orgname{Tohoku University}, \orgaddress{ \city{Sendai}, \postcode{980-8577}, \country{Japan}}}

\abstract{Altermagnets—collinear, zero–net-moment magnets with momentum-odd spin splitting protected by crystalline symmetries—offer a tunable route to suppress long-range antiferromagnetism while preserving strong short-range spin fluctuations.  We show that this environment robustly stabilizes unconventional superconductivity and naturally produces mixed-symmetry pairing. Through a strong-coupling analysis of a spin-anisotropic Hubbard model, we derive an anisotropic $t–J$ model where exchange interactions cooperatively enhance singlet $d$-wave and promote triplet $p$-wave pairing. Our mean-field analysis reveals a pairing evolution driven by altermagnetic anisotropy: for small spin anisotropy, the $d$-wave channel is enhanced, closely resembling the dominant pairing symmetry in cuprate superconductors, which suggests that weak spin anisotropies may be an essential ingredient in realistic models of these materials. Constrained-path quantum Monte Carlo simulations confirm this picture, showing a regime where dominant $d$-wave correlations coexist with an emergent $p$-wave component near optimal doping. As spin anisotropy increases, strong $C_2$ anisotropy and spin splitting activate the triplet channel, leading to a stable $d+p$ mixed-pairing state. This synergistic state exhibits significantly enhanced overall pairing strength, predicting a higher superconducting transition temperature. }




\maketitle
 Unconventional superconductivity emerging from strongly correlated, net-zero-magnetization backgrounds remains one of the central puzzles in condensed matter physics~\cite{Scalapino2012-jr}. In cuprate and heavy-fermion systems, long-range antiferromagnetism is suppressed by doping, while short-range spin fluctuations persist and are believed to mediate $d$-wave pairing~\cite{Anderson1987-cv}.  Since the discovery of cuprates with predominant $d_{x^2-y^2}$ pairing~\cite{Bednorz1986,Tsuei2000-fb,RevModPhys.67.515}, the two-dimensional one-band Hubbard and $t$-$J$ models have served as minimal frameworks to examine how strong local repulsion and magnetic fluctuations can generate effective pairing glue~\cite{Imada1998-fh,Keimer2015-jp,Jiang2021-hj,Qin2022-ad}. However, doping alone may not sufficiently account for unconventional superconductivity over broad ranges, motivating the identification of complementary ingredients that cooperate with doping.

Notably, a striking spin anisotropy in cuprates foreshadows the concept of altermagnetism, where multi-orbital studies have shown that oxygen moments in  CuO$_2$ planes can generate a $d$-wave-like spin pattern with momentum-dependent splitting yet zero net magnetization~\cite{PhysRevX.12.040501, songAltermagnetsNewClass2025a,doi:10.1126/sciadv.aaz8809,doi:10.1126/science.aao0980}. This phenomenon finds its formalization in altermagnetism, which has recently emerged as a distinct class of magnetism characterized by compensating collinear spin structures related to crystal rotational symmetries~
\cite{PhysRevX.12.031042,baiAltermagnetismExploringNew2024a, rimmlerNoncollinearAntiferromagneticSpintronics2024a,  zhangCrystalsymmetrypairedSpinValley2025}. 
Combining key features of ferromagnets and antiferromagnets, altermagnets exhibit momentum-dependent spin splitting while preserving symmetry-enforced zero net magnetization~\cite{PhysRevX.12.040501,baiAltermagnetismExploringNew2024a,https://doi.org/10.1002/advs.202503235}. Intense theoretical and experimental efforts have led to the identification of several altermagnetic materials, including $\text{Ru}\text{O}_2$~\cite{PhysRevLett.122.017202, doi:10.1126/sciadv.adj4883}, $\text{MnTe}$~\cite{Amin2024}, $\text{Mn}_5\text{Si}_3$~\cite{Surgers2024}, $\text{CrSb}$~\cite{reimers2024natcomm, Zhou2025}, 
and $\text{KV}_2\text{Se}_2\text{O}$~\cite{Jiang2024-tm}. Depending on the underlying crystal symmetry, altermagnets realize distinct spin-splitting textures. These give rise to a spectrum of novel transport and optical phenomena, establishing altermagnetism as a versatile platform for exploring correlated and nonequilibrium spin physics.

Recent theoretical works predict unconventional pairing phenomena in altermagnetic systems, including finite-momentum Cooper pairing~\cite{zhangFinitemomentumCooperPairing2024a}, superconducting diode effect \cite{cv8s-tk4c}, and topological superconductivity induced by altermagnetic spin splitting~\cite{GhorashiPRLAlt2024,dlpb-gfct}. However, the cooperative influence of altermagnetic hopping anisotropy $t_A$ and doping $n$ on the competition between magnetic order and pairing in the mean-field framework and beyond remains largely unexplored. In this work, we study a minimal Fermi-Hubbard Hamiltonian with spin-selective, anisotropic nearest-neighbor hopping parameterized by $t_A$. Starting from half-filling ($n=1$), we vary doping $\delta=|1-n|$ to introduce mobile carriers and examine its interplay with $t_A$.

\begin{figure}[h]
\centering 
\includegraphics[width=0.9\textwidth]{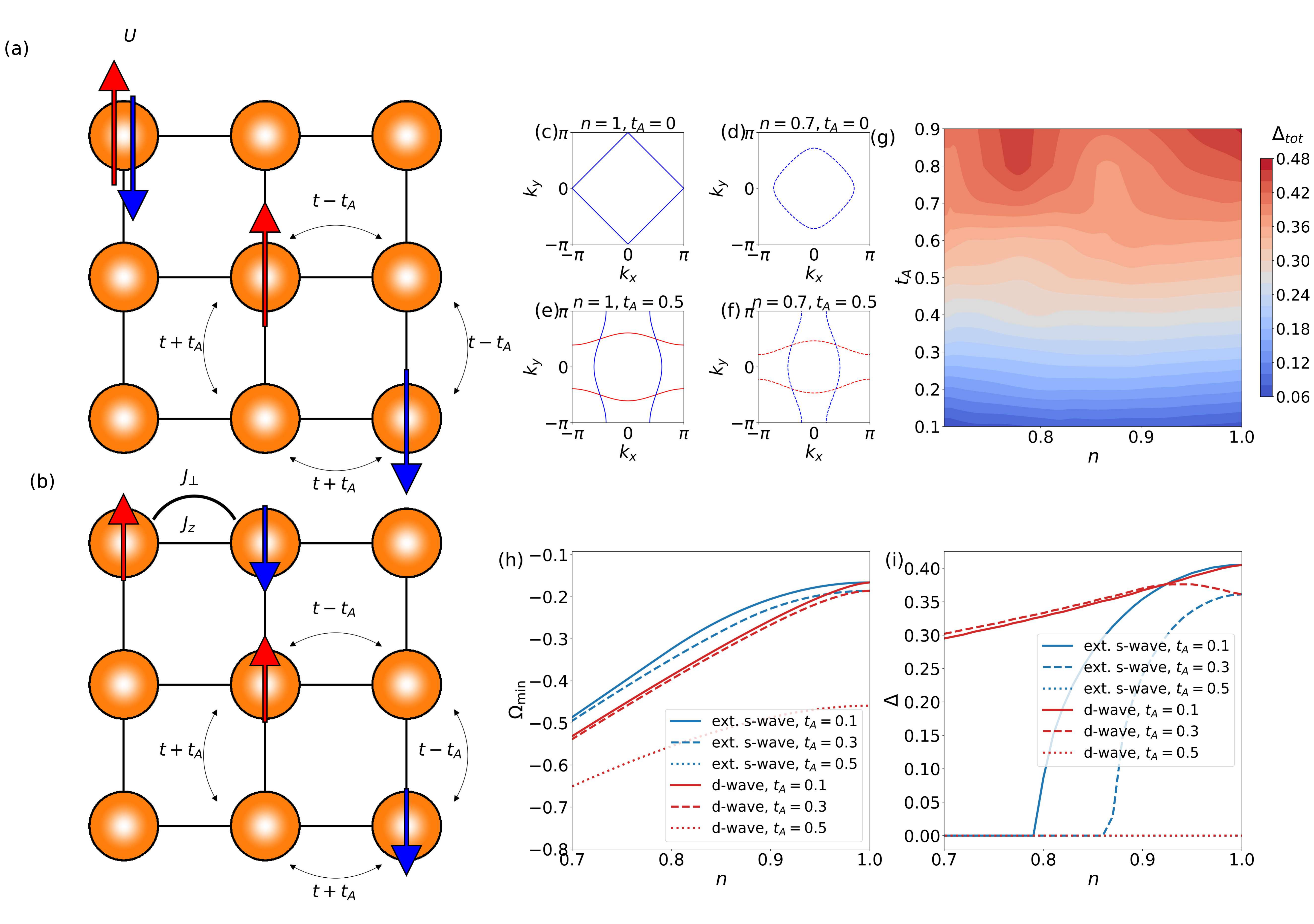}
\caption{\textbf{Spin-anisotropic Hubbard model, mapping to the $t$--$J$ model, and pairing analysis.} 
(a) Schematic of the Hubbard model with spin-anisotropic nearest-neighbor hopping $t \pm t_A$ 
under on-site repulsion $U$. 
(b) In the strong-coupling limit, the system maps onto an anisotropic $t$--$J$ Hamiltonian 
characterized by $J_z$ and $J_\perp$. 
(c--f) Representative Fermi surfaces at different fillings $n$ and anisotropies $t_A$, 
illustrating the reconstruction from a perfectly nested square to distorted and spin-split contours.
(g) QMC evaluation of the total pairing strength $\Delta_{\mathrm{tot}}$ in the $(n,t_A)$ 
parameter space, showing a broad enhancement with increasing anisotropy. 
(h,i) Comparison of extended $s$-wave (blue) and $d$-wave (red) states within the $t$--$J$ model. 
Although the nominal grand potential $\Omega_{\min}$ of the $s$-wave branch is slightly lower (h), 
its gap amplitude remains vanishingly small across the entire range (i), corresponding to a normal state. 
In contrast, the $d$-wave solution develops a robust finite gap whose onset filling decreases 
and magnitude grows with $t_A$, establishing $d$-wave as the dominant superconducting instability 
stabilized by spin-anisotropic hopping in the altermagnetic background. }\label{fig1}
\end{figure}

\subsubsection*{Unconventional pairing in $t-J$ model with doping and anisotropy}

We introduce a spin-dependent anisotropic hopping that physically motivated as an effective model representing materials where a high-energy magnetic order, such as antiferromagnetism, creates a complex, spin-dependent energy landscape for electrons~\cite{PhysRevB.110.144412,PhysRevLett.112.017205}. Because this magnetic energy scale is often orders of magnitude larger than that of superconductivity, its primary effect can be captured by an effective spin-dependent hopping parameter; therefore, we consider the square-lattice Fermi–Hubbard Hamiltonian with spin-dependent anisotropic hopping~\cite{li2025enhancement,RevModPhys.78.17},

\[H =-\!\!\sum_{i,\sigma}\! \Bigl( t_{\hat{x},\sigma}\,c_{i,\sigma}^{\dagger}c_{i+\hat{x},\sigma} +t_{\hat{y},\sigma}\,c_{i,\sigma}^{\dagger}c_{i+\hat{y},\sigma} +\text{H.c.} \Bigr) +U\sum_{i}n_{i,\uparrow}n_{i,\downarrow}, \]
where \(c^{\dagger}_{i,\sigma}\) (\(c_{i,\sigma}\)) creates (annihilates) a fermion with spin
\(\sigma=\uparrow,\downarrow\) at site \(i\) and \(n_{i,\sigma}=c^{\dagger}_{i,\sigma}c_{i,\sigma}\). A schematic illustration of the model on the 2D square lattice is shown in Fig.~\ref{fig1}(a).
The anisotropy is encoded by a single tunable parameter \(t_A\in[0,1]\):
\(
t_{\hat{y}\downarrow}=t_{\hat{x}\uparrow}=t-t_A,\quad
t_{\hat{x}\downarrow}=t_{\hat{y}\uparrow}=t+t_A.
\) Because \(t_{l,\uparrow}+t_{l,\downarrow}=2t\), the total
bandwidth is conserved and the spin populations remain balanced with
\(\langle n_{i,\uparrow}\rangle=\langle n_{i,\downarrow}\rangle=n/2\). Thus \(t_{A}=0\) reproduces the isotropic Hubbard model, whereas \(t_{A}=1\) corresponds to the extreme
\(C_{2}\) limit in which each spin species hops strictly along one
lattice axis. The anisotropy generates an altermagnetic spin splitting that is odd in momentum and averages to zero, ensuring zero net magnetization.

In the strong coupling regime ($U \gg t$), we perform a Schrieffer–Wolff transformation~\cite{PhysRev.149.491,BRAVYI20112793} to integrate out doubly occupied states, yielding an effective anisotropic $t$--$J$ model [see Fig.~\ref{fig1}(b)]:

\[ \begin{split} H_{t-J} = -\sum_{\langle ij \rangle,\sigma} t_{ij,\sigma}\, \tilde{c}^\dagger_{i\sigma} \tilde{c}_{j\sigma} + \sum_{\langle ij \rangle} \Bigl( J_\perp \mathbf{S}_i^\perp \!\cdot \mathbf{S}_j^\perp + J_z S^z_i S^z_j - J_z \frac{n_i n_j}{4} \Bigr). \end{split}\]
with $\tilde{c}_{i\sigma} = c_{i\sigma} (1 - n_{i\bar{\sigma}})$ the projected fermion operator, and anisotropic superexchange couplings $J_\perp = 4(t^2 - t_A^2)/U$ and $J_z = 4(t^2 + t_A^2)/U$.
Intuitively, the direction in which a given spin hops more easily produces stronger associated superexchange. This yields two effective couplings: a component that grows with $t_A$ and favors spin-singlet channels (extended-$s$ and $d$-wave), and an anisotropic component that promotes spin-triplet $p$-wave tendencies.  Upon doping, mobile carriers traverse this anisotropic spin background; because only $C_2$ symmetry remains, a $p$-wave component is symmetry-allowed to admix with the leading $d$-wave state, producing mixed-symmetry pairing whose relative weights evolve with $t_A$ and carrier concentration.

Within the $t$--$J$ analysis, we benchmark the grand potential $\Omega_{\min}$ of the extended-$s$ and $d$-wave solutions as a function of filling $n$ and anisotropy $t_A$. As shown in Fig.\ref{fig1}(h), the $d$-wave branch consistently attains a lower $\Omega_{\min}$ across the entire density range, indicating a robust stabilization of $d$-wave pairing by the altermagnetic anisotropy. This enhancement is rooted in the emergent altermagnetic background. Figure \ref{fig1}(g) shows that the total momentum-domain spin polarization, $\Delta_{\text{tot}}$, increases with $t_A$ for $U=4$, signifying a stronger spin splitting. This environment, visualized by the spin-split Fermi surfaces in Fig. \ref{fig1}(c-f), at the same time, the nesting condition is disrupted, resulting in destruction of antiferromagnetic instability at isotropic half filling as well as creating favorable conditions for superconducting fluctuations. 
Our prior work~\cite{li2025enhancement} demonstrated that altermagnetic spin splitting enhances $d$-wave pairing via a resonating valence bond~(RVB)-like mechanism at half-filling~\cite{10.1126/science.235.4793.1196,BASKARAN1987973}, where Fermi-surface reconstruction suppresses antiferromagnetism and promotes short-range singlet and triplet correlations. 
Here, we extend this mechanism to the doped regime. Doping introduces mobile carriers that convert the short-range correlations into coherent mixed singlet and triplet pairs. Combined with altermagnetic anisotropy, the cooperative effect further enhances superconductivity beyond the half-filling. This generalizes the application of RVB-like mechanism to a richer prospects and broader regime, reinforcing the unconventional pairings in the altermagnetic system.
 
As $t_A$ increases from $0.1$ to $0.3$, the $d$-wave gap $\Delta_d$ is significantly larger than the extended $s$-wave gap $\Delta_s$, indicating that $d$-wave pairing stably dominates in this parameter range. However, when $t_A$ further increases to $0.5$, both $\Delta_d$ and $\Delta_s$ drop sharply and approach zero, reflecting strong suppression of both pairing channels under large $t_A$, in which $p$-wave pairing plays the main role where we will discussed later. This non-monotonic behavior demonstrates that although $d$-wave pairing holds a clear advantage over extended $s$-wave at moderate $t_A$, its amplitude does not increase continuously with $t_A$.  In the presence of only $C_2$ symmetry, doping-induced carriers may promote mixing between the spin-singlet $d$-wave and spin-triplet $p$-wave component, leading to mixed-symmetry pairing. The evolution of $\Delta_d$ and $\Delta_s$ with $t_A$—first maintaining dominance and then being jointly suppressed—illustrates a general trend of overall superconducting instability in the system as it enters the high-$t_A$ regime.

\subsubsection*{Competing density waves restricting pairing enhanced regimes }\label{sec1}

\begin{figure}[h]
\centering
\includegraphics[width=1\textwidth]{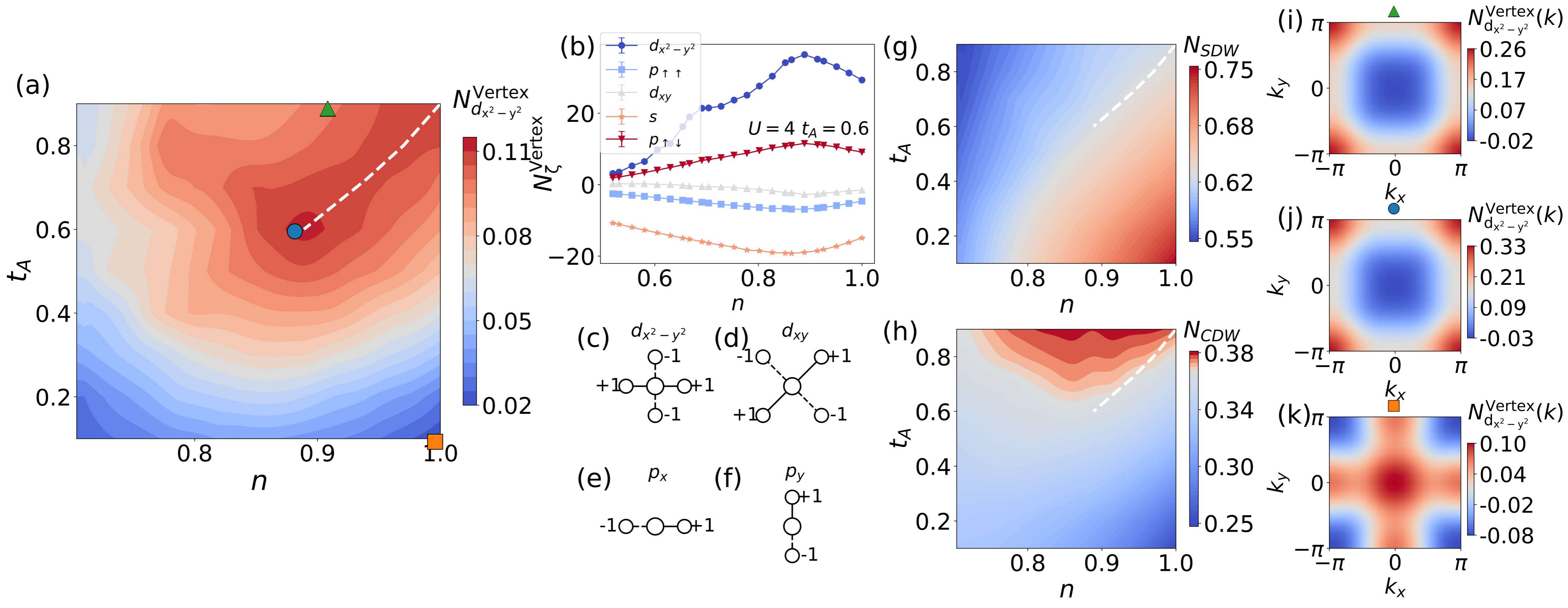}
\caption{\textbf{Enhancement of $d$-wave pairing restricted by  competing density-wave channels.} 
(a) Quantum Monte Carlo evaluation of the total vertex $d$-wave pairing function $N^{\mathrm{Vertex}}_{d_{x^2-y^2}}$ 
in the $(n,t_A)$ parameter space, showing a pronounced maximum around 
$n \approx 0.88$ and $t_A \approx 0.6$ (blue circle), with pairing strength 
increasing along the dashed trajectory. 
(b) Filling dependence of vertex functions for multiple pairing channels, 
demonstrating the dominance of $d_{x^2-y^2}$ and $p_{\uparrow\downarrow}$ over others. 
(c--f) Form factor $\eta^\zeta$ of the $d_{x^2-y^2},\ d_{xy},\ p\ (p_x$ or $p_y$) pairing symmetry. 
(g,h) the spin structure factor $N_{\mathrm{SDW}}^{local}$ and the charge structure factor $N^{local}_{\mathrm{CDW}}$ in the $(n,t_A)$ space. 
(i--k) Momentum-resolved vertex function for the $d_{x^2-y^2}$-wave channel at the representative parameter points indicated by the blue markers.
Together, these results establish that $d$-wave pairing is strongly enhanced 
in the region where both SDW and CDW tendencies are weakened, consistent with 
the RVB picture in which short-range spin fluctuations mediate superconductivity 
once density-wave instabilities are suppressed.}\label{fig2}
\end{figure}

Taken together, the $t$--$J$ analysis shows that an altermagnetic background
can stabilize and strengthen the $d$-wave superconducting solution by lowering
its onset filling and deepening the minimum in the free-energy landscape
$\Omega(\Delta_d,\Delta_p)$. Although spin density wave~(SDW) and  charge density wave~(CDW) channels are not treated
explicitly at this level, the $t$--$J$ results suggest that, in the full Hubbard
model, $d$-wave superconductivity should be optimized in the parameter regime
where spin-anisotropy–induced altermagnetic correlations reduce the tendency
toward density-wave ordering~\cite{doi:10.1126/science.aat4708}. In the following, we test this expectation in the
underlying Hubbard model using large-scale qauntum Monte Carlo~(QMC) simulations, which quantify
the evolution of $d$-wave pairing and SDW/CDW correlations on an equal footing. Specifically, we use constraint path QMC~\cite{Zhang_1995, Zhang1997-ry}, in which the constraint path approximation is implemented to mediate the sign problem.
As shown in Fig.~\ref{fig2}(a), the $d_{x^2-y^2}$ channel exhibits a pronounced 
peak in the vertex pairing function $N^{\text{Vertex}}_{d_{x^2-y^2}}$ within the 
$(n,t_A)$ parameter space.

\[
N^{\rm Vertex}_{d_{x^2-y^2}}(\mathbf{k}) = \frac{1}{N}\sum_{i,j} \exp\left[i\mathbf{k}\cdot(\mathbf{r}_i - \mathbf{r}_j)\right] C^{\rm Vertex}_{d_{x^2-y^2}}(i,j),
\]
where \(
C^{\rm Vertex}_{d_{x^2-y^2}}(i,j) =
\sum_{\delta_\zeta,\delta'_\zeta} \Bigl(
   \langle \Delta_{d_{x^2-y^2}}^\dagger(i)\,
           \Delta_{d_{x^2-y^2}}(j) \rangle
   - \eta_{\delta_\zeta}^{d_{x^2-y^2}} \eta_{\delta'_\zeta}^{d_{x^2-y^2}}
     G^{\uparrow}_{i,j} G^{\downarrow}_{i+\delta_\zeta,j+\delta'_\zeta}
)
\)
with $\Delta_{d_{x^2-y^2}}^\dagger(i) = c_{i\uparrow}^\dagger(c_{i+x\downarrow}^\dagger - c_{i+y\downarrow}^\dagger + c_{i-x\downarrow}^\dagger - c_{i-y\downarrow}^\dagger)$ with $\delta_\zeta^{(')}$ denoting nearest-neighbor bonds. The distinct form factors
$\eta^\zeta$ 
of these pairing symmetries on the square lattice are illustrated in Fig.~\ref{fig2}(c-f).

The restricted regime of enhancement suggests that competing orders may suppress the unconventional pairing, so we analyse the density-wave channels using QMC. The longitudinal spin structure factor $N_{S_z}(\mathbf{k}) = \frac{1}{N}\sum_{i,j} e^{i\mathbf{k}\cdot(\mathbf{r}_i-\mathbf{r}_j)}
  \langle S_i^{z} S_j^{z}\rangle$,where $
  S_i^{z} = \frac{1}{2}(n_{i\uparrow}-n_{i\downarrow})$,
is used to construct the  spin structure factor measure
 $ N_{\mathrm{SDW}} = \frac{1}{L^2}\sum_{\mathbf{k}} N_{S_z}(\mathbf{k})$,whose distribution in the $(n,t_A)$ plane is shown in Fig.~\ref{fig2}(g). 
$N_{\mathrm{SDW}}$ is largest close to half filling and decreases when moving away from $n\simeq 1$, indicating that strong antiferromagnetic correlations are confined to the vicinity of half filling and are weakened along the dashed trajectory where the $d$-wave vertex is enhanced. 
Within an RVB picture, the suppression of long-range N\'eel order leaves robust short-range spin fluctuations that act as the glue for sign-changing pairing, naturally accounting for the growth of the $d$-wave vertex as $N_{\mathrm{SDW}}$ is reduced.

Charge order constitutes a second competitor. 
Its correlations are characterised by
 $ N_{\mathrm{CDW}}(\mathbf{k}) = \frac{1}{N}\sum_{i,j} e^{i\mathbf{k}\cdot(\mathbf{r}_i-\mathbf{r}_j)}
  \langle n_i n_j\rangle,$ and the charge structure factor $N_{\mathrm{CDW}} = \frac{1}{L^2}\sum_{\mathbf{k}} N_{\mathrm{CDW}}(\mathbf{k})$
is displayed in Fig.~\ref{fig2}(h). 
Away from half filling, around $n\!\approx\!0.88$ where the $d$-wave vertex attains its maximum, $N_{\mathrm{CDW}}$ shows a pronounced ridge at small $t_A$, signalling enhanced CDW tendencies in this doping range. 
Along the dashed trajectory, increasing $t_A$ systematically suppresses $N_{\mathrm{CDW}}$, demonstrating that spin anisotropy weakens charge-order tendencies precisely in the density regime relevant for the $d$-wave peak. 
Phenomenologically, an enhanced $N_{\mathrm{CDW}}$ corresponds to an incipient charge modulation that would reconstruct the Fermi surface and deplete low-energy carriers that could otherwise participate in $d$-wave pairing \cite{Yu2021}. 
A moderate spin-anisotropic hopping $t_A$ therefore weakens both SDW near $n\!\approx\!1$ and CDW near $n\!\approx\!0.88$, preventing such depletion and allowing the $d$-wave vertex to reach its maximum. From the QMC result with medium range of $U$, we noted that the $d$-wave pairing becomes the dominant pairing channel in the regime of finite doping around $n\approx 0.85$ and strong anisotropy $t_A=0.7-0.9$~(See supplementary material Sec.~S6).
This is consistent with the $t$--$J$ model analysis in the strong $U$ limit, where the anisotropy lowers the onset doping for pairing and sustains a finite $d$-wave gap $\Delta_d$.

\subsubsection*{Prospective mixed $p$- and $d$-wave superconductor with finite doping and anisotropy}

\begin{figure}[h]
\centering
     \includegraphics[width=0.9\textwidth]{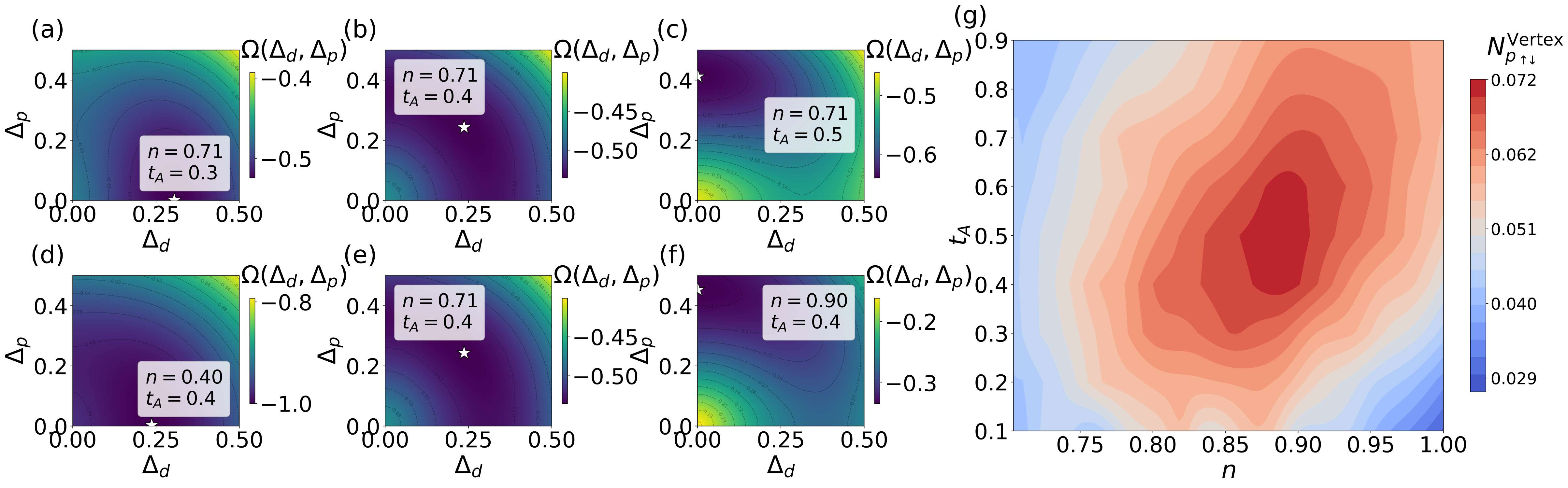}
\caption{\textbf{Enhancement of $p$-wave pairing and emergence of $d+p$ coexistence.}  
(a--c) Free-energy landscape $\Omega(\Delta_d,\Delta_p)$ of the $t$--$J$ model for representative anisotropies $t_A=0.3,0.4,0.5$: the global minimum evolves from a pure $d$-wave state to a $d+p$ mixed state (star symbol in panel b) and finally towards a regime dominated by the $p$-wave component. 
(d--f) Free-energy landscape $\Omega(\Delta_d,\Delta_p)$ of the $t$--$J$ model for representative anisotropies $n=0.4,0.71,0.9$: the global minimum evolves from a pure $d$-wave state to a $d+p$ mixed state (star symbol in panel b) and finally towards a regime dominated by the $p$-wave component.(g) Quantum Monte Carlo results for the vertex pairing function 
$N^{\mathrm{Vertex}}_{p_{\uparrow\downarrow}}$ in the $(n,t_A)$ parameter space, showing a pronounced enhancement around $n \approx 0.85$--$0.90$ and $t_A \approx 0.45$--$0.55$.
Taken together, QMC and $t$--$J$ mean-field results demonstrate that near optimal filling and intermediate anisotropy, $p$-wave pairing is strongly enhanced and cooperates with $d$-wave order, stabilizing a mixed-symmetry superconducting state.
}\label{fig3}
\end{figure}
Within the $t$--$J$ model, the mean-field free-energy landscapes $\Omega(\Delta_d,\Delta_p)$ 
shown in Fig.~\ref{fig3}(a--f) reveal how anisotropy $t_A$ and filling $n$ jointly reshape 
the competition between the $d$- and $p$-wave pairing channels. 
Figures~\ref{fig3}(a--c), obtained at fixed filling $n=0.71$ for 
$t_A = 0.3,\,0.4,\,0.5$, illustrate a clear evolution of the global minimum: 
at weak anisotropy the minimum corresponds to a pure $d$-wave state; 
at intermediate anisotropy [Fig.~\ref{fig3}(b)] it shifts to a mixed $d{+}p$ configuration 
(marked by the star symbol); and at stronger anisotropy it gradually moves toward a regime 
dominated by the $p$-wave component. 
This evolution originates from the symmetry reduction $C_4 \!\to\! C_2$, which weakens 
$(\pi,\pi)$ nesting in the singlet sector, while the anisotropic hopping simultaneously 
produces momentum-dependent spin splitting that amplifies the triplet kernel.

A complementary trend is found in Fig.~\ref{fig3}(d--f), where the anisotropy is 
fixed at $t_A = 0.4$ and the filling varies over $n = 0.40,\,0.71,\,0.90$. 
At low filling, the free-energy minimum remains predominantly $d$-wave; 
near $n=0.71$, a stable mixed $d{+}p$ minimum emerges; 
and at higher filling the landscape shifts toward $p$-wave dominance. 
In the $d{+}p$ coexistence region, the superconducting order parameter is intrinsically 
multi-component, and the relative phase between the singlet and triplet components acquires 
a finite stiffness enabled by the altermagnetic anisotropy. 
This generates a neutral Leggett mode—an out-of-phase oscillation between the two condensates—
with a finite gap. The Leggett excitation provides a characteristic dynamical fingerprint of 
mixed-symmetry superconductivity and should appear as a low-energy resonance in Raman $A_{1g}$ 
or THz phase spectroscopy~\cite{Cuozzo2024,PhysRevB.102.144519,Giorgianni2019}.

Fig.~\ref{fig3}(g) provides a complementary non-perturbative QMC perspective. 
The vertex pairing strength $N^{\mathrm{Vertex}}_{p_{\uparrow\downarrow}}$ develops a pronounced 
ridge around $t_A\!\sim\!0.4$--0.6 and $n\!\sim\!0.85$--$0.90$, indicating that moderate 
anisotropy and near-optimal filling substantially enhance the spin-triplet channel. 
Taken together with the mean-field landscapes, this shows that the $d{+}p$ coexistence region 
is stabilized precisely where anisotropy-enhanced triplet pairing intersects with the 
singlet $d$-wave pairing, and where the Leggett mode is expected to be most robust~\cite{PhysRevB.105.094520,Cuozzo2024}.

In the weak-anisotropy limit, our calculations show that the $d$-wave 
pairing channel is significantly strengthened. The resulting gap structure 
closely resembles that of the cuprate superconductors, where a robust 
$d$-wave superconducting state emerges from strong electronic correlations. 
Notably, the fact that the $d$-wave response remains dominant even when a 
small altermagnetic anisotropy is present suggests that a realistic 
description of cuprates may also require accounting for weak spin 
anisotropies that are usually neglected in minimal models. The presence of 
a moderate spin anisotropy can act in synergy with the $d$-wave symmetry, 
leading to a further enhancement of superconductivity in doping regimes 
where it is already promoted. Beyond solid-state cuprates, the same 
spin-anisotropic mechanism can be implemented in optical-lattice setups via 
spin-dependent tunneling and engineered time-reversal-symmetry breaking \cite{Hart2015}. 
This approach, which isolates the role of time-reversal symmetry breaking 
inherent to altermagnetism, is further justified by its direct and tunable 
realization in modern cold-atom experiments~\cite{PhysRevLett.111.185301,doi:10.1126/science.1201351,jau2016entangling,zeiher2016many}.

As the spin-anisotropic hopping $t_A$ increases, the system gradually 
departs from the cuprate-like regime and enters a domain where the momentum-dependent spin splitting begin to 
activate the triplet pairing channels. Our self-consistent solutions show 
that beyond a critical anisotropy scale, the gap function develops a finite 
$p$-wave component coexisting with the dominant $d$-wave structure, forming 
a stable mixed $d{+}p$ superconducting state. Importantly, the enhanced 
spin splitting associated with large $t_A$ strengthens the overall pairing 
kernel, leading to a larger effective pairing amplitude than in the 
small-$t_A$ $d$-wave regime. This synergistic $d{+}p$ state, with its 
reinforced pairing strength, may lead to an elevated superconducting 
transition temperature. Our results thus indicate that altermagnetic spin 
anisotropy provides a natural mechanism not only for generating 
triplet-enriched superconductivity but also for enhancing $T_c$ in the 
intermediate to large $t_A$ regime, offering a new avenue for designing 
high-temperature superconductors~\cite{xuFrustratedAltermagnetismCharge2023a}.

\section*{Methods}\label{sec11}
\subsection*{Mean-field approach in altermagnetic $t-J$ model}
\label{sec3}
In the strong-coupling limit ($U \gg t$), the spin-dependent anisotropic hopping leads to an effective anisotropic $t$–$J$ model, expressed as
\[
H=-\sum_{i, \sigma, l=\hat{x}, \hat{y}}\left(t_{l, \sigma} C_{i, \sigma}^{\dagger} C_{i+l, \sigma}+h . c .\right)+\sum_{\langle i, j\rangle} (J_{\perp} \mathbf{S}_i^{\perp} \cdot \mathbf{S}_j^{\perp}+J_z \mathbf{S}_i^Z \cdot \mathbf{S}_j^Z-J_Z \frac{n_i n_j}{4}).
\]
Here, $J_{\perp}=J-J_A$ and $J_z=J+J_A$ with $J=\frac{4 t^2}{U}$ and $J_A=\frac{4 t_A^2}{U}$.

In the mean field framework based on the slave boson method, the Hamiltonian can be written as:

\begin{align}
H_{\mathrm{eff}} = 
&\sum_{\mathbf{k}\sigma} \varepsilon_{\mathbf{k}\sigma} C_{\mathbf{k}\sigma}^\dagger C_{\mathbf{k}\sigma}
- J_z \sum_\mathbf{k} \left( C_{\mathbf{k}\uparrow}^\dagger C_{-\mathbf{k}\downarrow}^\dagger \Delta_\eta(\mathbf{k}) + \Delta_\eta^*(\mathbf{k}) C_{-\mathbf{k}\downarrow} C_{\mathbf{k}\uparrow} \right) \nonumber\\
&- 4J_A \sum_\mathbf{k} \left( \Delta_p(\mathbf{k}) C_{\mathbf{k}\uparrow}^\dagger C_{-\mathbf{k}\downarrow}^\dagger  + \Delta_p ^*(\mathbf{k}) C_{-\mathbf{k}\downarrow} C_{\mathbf{k}\uparrow} \right)
+ J_z  |\Delta_\eta|^2 + 8J_A  |\Delta_p|^2.\nonumber
\end{align}
Here, \(\epsilon_{\mathbf{k}}=-2t(\cos k_x+\cos k_y) -\mu\) describes kinetic energy, \(\epsilon_A=-2t_A(\cos k_x - \cos k_y)\) encodes the altermagnetic splitting. 
We consider the extended \(s\)-wave (\(\Delta_s(\mathbf{k})=\Delta_s(\cos k_x+\cos k_y)\)), \(d\)-wave (\(\Delta_d(\mathbf{k})=\Delta_d(\cos k_x-\cos k_y)\))  and \(p\)-wave (\(\Delta_p(\mathbf{k})=\Delta_p(\sin k_x+i\sin k_y)\)) pairing, with the pairing amplitude \(\Delta_{s/p/d}\). 

By introducing the Nambu spinor \(\psi_{\mathbf{k}}=(C_{\mathbf{k}\uparrow},C_{\mathbf{k}\downarrow},C_{-\mathbf{k}\uparrow}^\dagger,C_{-\mathbf{k}\downarrow}^\dagger)^T\), the Hamiltonian can be written as:  

\begin{equation}
H_{\mathrm{eff}} =  \tfrac{1}{2} \sum_{\mathbf{k}} \psi_{\mathbf{k}}^{\dagger} \mathcal{H}(\mathbf{k}) \, \psi_{\mathbf{k}} 
+ \sum_{\mathbf{k}} \epsilon_{\mathbf{k}} + J_z N |\Delta_\eta|^2 + 8J_A N |\Delta_p|^2,\nonumber
\end{equation}
where the Bogoliubov–de Gennes (BdG) Hamiltonian takes the specific form:
\begin{equation}
\mathcal{H} (\mathbf{k}) = \epsilon_k \tau_z + \epsilon_A \sigma_z \tau_z - J_z \Delta_\eta (\mathbf{k}) \sigma_y \tau_y 
+ 8J_A \Delta_p (\sin k_x \sigma_x \tau_x - \sin k_y  \sigma_x \tau_y).\nonumber
\end{equation}
At zero temperature, the grand potential takes the form
\[
\Omega
=\sum_{\mathbf{k}}\epsilon_{\mathbf{k}}
-\tfrac12\sum_{\mathbf{k}}\bigl(E_{\mathbf{k}}^+ + E_{\mathbf{k}}^-\bigr)
+ J_z N |\Delta_\eta|^2 
+ 8J_A N |\Delta_p|^2,
\] 
where $E_{\mathbf{k}}^\pm$ denote the quasiparticle excitation spectra of the BdG Hamiltonian. From the underlying $t$–$J$ model, the exchange $J_Z$ grows with $t_A$ and favors spin-singlet channels ($d$-wave or extended $s$-wave), while the anisotropic exchange $J_A$ drives spin-triplet $p$-wave pairing.

\subsection*{Constraint path quantum Monte Carlo}

\label{sec6}
The constraint path quantum Monte Carlo~(CPQMC) method is a quantum Monte Carlo method implemented with a constraint path approximation~\cite{Zhang1997-ry}, mediating the sign problem encountered in QMC studies~\cite{PhysRevX.5.041041}. In CPQMC, the ground state wave function $\psi^{(n)}$ is represented by a finite ensemble of Slater determinants, i.e.,  
$\left|\psi^{(n)}\right\rangle \propto \sum_k\left|\phi_k^{(n)}\right\rangle$,
where $k$ is the index of the Slater determinants, and $n$ is the number of iteration. The overall normalization factor of the wave function has been omitted here. The propagation of the Slater determinants dominates the computational time, as follows
\begin{equation*}
\left|\phi_k^{(n+1)}\right\rangle \leftarrow \int d \vec{x} P(\vec{x}) B(\vec{x})\left|\phi_k^{(n)}\right\rangle .
\label{propagation}
\end{equation*}
where $\vec{x}$ is the auxiliary-field configuration, that we select according to the probability distribution function $P(\vec{x})$. The propagation includes the matrix multiplication of the propagator $B(\vec{x})$ and $\phi_k^{(n)}$. 
After a series of equilibrium steps, the walkers are the Monte Carlo samples of the ground state wave function $\phi^{(0)}$ and ground-state properties can be measured. 

The random walk formulation suffers from the sign problem due to the fundamental symmetry between the fermion ground state $|\Psi_0\rangle$ and its negative counterpart $-|\Psi_0\rangle$. Mathematically, this problem originates from the structure of the ground state wave function in Slater determinant space. We can define a nodal surface, $\mathcal{N} = \{ |\phi\rangle \mid \langle \Psi_0|\phi\rangle = 0 \}$, which partitions the entire Slater determinant space into two symmetric half-spaces, with the ground state wave function $|\Psi_0\rangle$ existing only in one of them. From the projection principle perspective, once a walker $|\phi\rangle$ lands on the nodal surface $\mathcal{N}$, i.e., $\langle \Psi_0|\phi\rangle = 0$, its contribution to the ground state becomes exactly zero for all subsequent imaginary time evolution:
\[
\langle \Psi_0|\phi\rangle = 0 \Rightarrow \langle \Psi_0|e^{-\tau H}|\phi\rangle = 0, \quad \tau > 0.
\]
Therefore, in principle, restricting the random walk strictly to the ``correct" half-space defined by $\mathcal{N}$ would yield the exact ground state while completely avoiding the sign problem. However, the true ground state $|\Psi_0\rangle$ and its nodal surface $\mathcal{N}$ are unknown, which constitutes the crux of the problem. The core idea of CPQMC is to approximate $|\Psi_0\rangle$ using a known trial wave function $|\Psi_T\rangle$, thereby defining a trial nodal surface,
$\mathcal{N}_T = \{ |\phi\rangle \mid \langle \Psi_T|\phi\rangle = 0 \}.$ We use this trial nodal surface $\mathcal{N}_T$ as an approximation to the true nodal surface $\mathcal{N}$ and enforce the constraint at each step of the random walk for all walkers $|\phi_k^{(n)} \rangle$:
\[
\langle \Psi_T | \phi_k^{(n)} \rangle > 0.
\]
In practical Monte Carlo simulations, this constraint proves highly effective: after sampling the auxiliary field $\vec{x}$ according to $\hat{P}(\vec{x})$ and computing the propagated new walker $|\phi^{(n+1)}\rangle = B(\vec{x})|\phi^{(n)}\rangle$, we immediately calculate its overlap with the trial wave function $\langle \Psi_T | \phi^{(n+1)} \rangle$. If this value $\leq 0$, the walker is considered to have crossed the trial nodal surface, and we consequently set its weight to zero and discard it. Only walkers satisfying the constraint condition are retained for the next generation of evolution. This yields an approximate solution to the ground-state wave function, $\left|\psi_0^c\right\rangle=\Sigma_\phi|\phi\rangle$. This approximation becomes exact if we have the ground state wave function as our trial wave function $\left|\psi_T\right\rangle=\left|\psi_0\right\rangle $.

The distribution of random walkers represents the ground-state wave function $\left|\psi_0^c\right\rangle$ under the constrained path approximation. Various expectation values can then be computed from a population of these walkers and their weights. We use the effective scheme of back propagation for the physical quantities estimation, including the pairing and structure factors,
\begin{equation*}
\langle\mathcal{O}\rangle_{\mathrm{BP}}=\lim _{\tau \rightarrow \infty} \frac{\left\langle\psi_T \exp \left(-\tau H_c\right)\right| \mathcal{O}\left|\psi_0^c\right\rangle}{\left\langle\psi_T \exp \left(-\tau H_c\right) \mid \psi_0^c\right\rangle} .
\label{back propagation}
\end{equation*}
The details of hyper-parameters in CPQMC calculations are given as follows. The number of walkers is 1000, the number of blocks for relaxation is 10, the number of blocks for growth estimate is 3, the number of blocks after relaxation is 10, the number of steps in each block is 640, the Trotter step size is 0.01, the growth-control energy estimate is -50. To ensure the stability of the data, we set the number of steps per block to 640 and the Trotter step size to 0.01.

For the choice of trial wavefunction, we use the Hartree wavefucntion. We adopt a density-only mean-field decoupling and set the spin-flip averages to zero, \(\langle S_i^{+}\rangle=\langle S_i^{-}\rangle=0\).
The on-site Hubbard interaction then reduces to
$U\sum_in_{i\uparrow}n_{i\downarrow}=U\sum_i(n_{i\uparrow}\left\langle n_{i\downarrow}\right\rangle+n_{i\downarrow}\left\langle n_{i\uparrow}\right\rangle-\left\langle n_{i\uparrow}\right\rangle\left\langle n_{i\downarrow}\right\rangle).$
The resulting mean-field Hamiltonian separates into spin-resolved components:
\begin{equation*}
H_{\text{MF}} = -\sum_{i,\sigma,l} \left( t_{l,\sigma} c_{i,\sigma}^\dagger c_{i+l,\sigma} + \text{h.c.} \right) + U \sum_{i,\sigma} \left( \langle n_{i,\bar{\sigma}} \rangle n_{i,\sigma} - \frac{1}{2} \langle n_{i,\uparrow} \rangle \langle n_{i,\downarrow} \rangle \right),
\end{equation*}
where $\bar{\sigma}$ denotes the opposite spin of $\sigma$. The self-consistency condition requires that the input densities $\langle n_{i,\sigma} \rangle^{\text{in}}$ and the output densities $\langle n_{i,\sigma} \rangle^{\text{out}}$ calculated from the eigen states of $H_{\text{MF}}$ converge, typically satisfying $\max_i |\langle n_{i,\sigma} \rangle^{\text{out}} - \langle n_{i,\sigma} \rangle^{\text{in}}| < \epsilon$. The single-particle wavefunctions $\phi_{\sigma,k}(i)$ is obtained by diagonalizing $H_{\text{MF},\sigma}$. The wavefunction determines the electron densities $\langle n_{i,\sigma} \rangle = \sum_{k=1}^{N_\sigma} |\phi_{\sigma,k}(i)|^2$, where $N_\sigma$ is the number of electrons with spin $\sigma$. As an additional numerical check, CPQMC is benchmarked against exact diagonalization on a small cluster, where the \(d_{x^2-y^2}\) vertex pairing correlator is reproduced with high accuracy over a representative range of anisotropy \(t_A\) (see Supplemental material Sec.~S6).

\section*{Declarations}
\textbf{Funding} This work is supported by the Shenzhen Fundamental Research Program (Grant No. JCYJ20250604145655074), the National Natural Science Foundation of China (Grant No. 12204130),  Shenzhen Key Laboratory of Advanced Functional Carbon Materials Research and Comprehensive Application (Grant No. ZDSYS20220527171407017), and Natural Science Foundation of Jiangsu Province (Grant No. BK20231320).\\
\textbf{Competing interests} The authors declare no competing interests. \\
\textbf{Ethics approval} Not applicable.\\
\textbf{Consent to participate} All authors agreed with the content.\\
\textbf{Consent for publication} All gave explicit consent to submit.\\
\textbf{Availability of data and materials} The data that support the plots in this paper and other findings of this study are available from the corresponding authors on reasonable request.\\
\textbf{Code availability} The code that supports the plots in this paper is available from the corresponding authors on reasonable request.\\
\textbf{Authors' contributions} X.S.Y. and H.K.T. conceived and supervised the project. J.L., J.Y.L. and P.Z. performed the strong-coupling analysis and mean-field calculations under the guidance of X.S.Y.. J.L. and J.Y.L. perform the simulation, analyze, and visualize the data under the guidance of H.K.T.  The theoretical interpretation of the results carried out by X.S.Y. and H.K.T.. All authors contributed to the review and editing of the manuscript.\\

\bibliography{sn-bibliography}

\end{document}


\title{Supplemental Material for ``Altermagnetic–doping interplay as a route to enhanced d-wave pairing in the Hubbard model"}


\author{Ji Liu}
\thanks{These authors contributed equally.}
\affiliation{School of Science, Harbin Institute of Technology, Shenzhen, 518055, China}
\affiliation{Department of Physics, Jiangsu University, Zhenjiang, 212013, China.}
\affiliation{Shenzhen Key Laboratory of Advanced Functional Carbon Materials Research and Comprehensive Application, Shenzhen 518055, China.}

\author{Jianyu Li}
\thanks{These authors contributed equally.}
\affiliation{Institute for Materials Research,Tohoku University,Sendai,980-85775,Japan}

\author{Peng Zhang}
\affiliation{School of Science, Harbin Institute of Technology, Shenzhen, 518055, China}
\affiliation{Shenzhen Key Laboratory of Advanced Functional Carbon Materials Research and Comprehensive Application, Shenzhen 518055, China.}

\author{Xiaosen Yang}
\email{yangxs@ujs.edu.cn}
\affiliation{Department of Physics, Jiangsu University, Zhenjiang, 212013, China.}

\author{Ho-Kin Tang}
\email{denghaojian@hit.edu.cn}
\affiliation{School of Science, Harbin Institute of Technology, Shenzhen, 518055, China}
\affiliation{Shenzhen Key Laboratory of Advanced Functional Carbon Materials Research and Comprehensive Application, Shenzhen 518055, China.}

\date{\today} 
 \maketitle 
In this Supplemental Material, we present additional computational results and the details of the method we used.
Sec.~\ref{sec1} presents a mean-field derivation of the $t$--$J$ model with altermagnetic dispersion,
together with several additional results that extend the discussion in the main paper.
Sec.~\ref{sec2} reports finite-size scaling of the vertex pairing correlators, establishing that the
\(d_{x^2-y^2}\) enhancement persists in the thermodynamic limit.
Sec.~\ref{sec3} surveys additional pairing channels (on-site \(s\), \(p_{\uparrow\uparrow}\), and \(d_{xy}\))
and quantifies their subleading roles relative to \(d_{x^2-y^2}\).
Sec.~\ref{sec4} provides momentum-resolved vertex maps for the \(d_{x^2-y^2}\) and \(p_{\uparrow\downarrow}\) channels,
highlighting their complementary \(k\)-space structures.
Sec.~\ref{sec5} presents momentum-resolved SDW and CDW correlation functions, clarifying how competing density-wave
tendencies evolve along the representative trajectory in the main paper.
Sec.~\ref{sec6} analyzes the momentum-resolved spin polarization \(\Delta n(\mathbf{k})\) versus filling \(n\) and anisotropy \(t_A\),
clarifying how altermagnetic spin splitting evolves across the Brillouin zone.
Sec.~\ref{sec7} quantifies the competition between the leading singlet channels by examining
\(N_{d_{x^2-y^2}}/N_{s^{\mathrm{ext}}}\), demonstrating a robust doped window at large \(t_A\) where
\(d_{x^2-y^2}\) overtakes the extended-\(s\) channel.Finally, Sec.~\ref{sec8} benchmarks the constrained-path quantum Monte Carlo (CPQMC) results against exact diagonalization (ED)
on a small cluster, demonstrating that the \(d_{x^2-y^2}\) vertex pairing correlator is reproduced with high accuracy
over a representative range of anisotropy \(t_A\).

\section{Mean-field derivation of $t-J$ model with altermagnetic dispersion}
\label{sec1}
We show the detailed derivation of the $t-J$ model from the altermagnetic Hubbard model.
The altermagnetic Hubbard model with spin-dependent polarization and zero-net magnetism. The spin-dependent anisotropic Hubbard model on a 2D square lattice can be written as following: 
\begin{align}
H = H_t + H_U 
=-\sum_{\substack{<ij>, \sigma}}t_{ij,\sigma} c_{i,\sigma}^\dagger c_{j,\sigma}   + U \sum_i n_{i,\uparrow} n_{i,\downarrow},\nonumber
\end{align}
with spin-dependent anisotropic hopping $ t_{\hat{y}\downarrow} = t_{\hat{x}\uparrow} = t - t_A $ and $t_{\hat{x}\downarrow} = t_{\hat{y}\uparrow} = t + t_A$.

In the strong coupling limit ($ U \gg t$), doubly occupied states are energetically suppressed. We can define project operators to partition the Hilbert space into no double occupancy states ($\mathcal{P}$) and states with double occupancy ($\mathcal{Q}$) with $\mathcal{P}+\mathcal{Q}=1$, 
$\mathcal{P}=\prod_{i=1}^{N}(1-\nu_{i})$, and $\nu_{i} = n_{i,\uparrow}n_{i,\downarrow}$.  For the Hubbard term, we have
\begin{align}
\mathcal{P}H_{U}\mathcal{P}=&\mathcal{P}H_{U}\mathcal{Q}= \mathcal{Q}H_{U}\mathcal{P}=0,  \nonumber\\
\mathcal{Q}H_{U}\mathcal{Q}=&H_{U}. \nonumber
\end{align}

Then,
\begin{align}
H=(\mathcal{P}+\mathcal{Q}) H (\mathcal{P}+\mathcal{Q})= (\mathcal{P}+\mathcal{Q}) H_{t} (\mathcal{P}+\mathcal{Q}) +H_{U}.\nonumber
\end{align}

We have a trivial identity $c_{i \sigma} \equiv c_{i \sigma}\left[\left(1-n_{i \bar{\sigma}}\right)+n_{i \bar{\sigma}}\right]$. Thus, we can rewrite the hopping term $H_{t}=T_{h}+T_{d}+T_{mix}$ with
\begin{align}
T_{h}=&-\sum_{\substack{<ij>, \sigma}} t_{ij,\sigma}\left(1-n_{i, \bar{\sigma}}\right) c_{i, \sigma}^{\dagger} c_{j, \sigma}\left(1-n_{j, \bar{\sigma}}\right) ,  \nonumber\\
T_{d}=&-\sum_{\substack{<ij>, \sigma}} t_{ij,\sigma} n_{i, \bar{\sigma}} c_{i, \sigma}^{\dagger} c_{j, \sigma} n_{j, \bar{\sigma}},   \nonumber\\
T_{\mathrm{mix}}=&-\sum_{\substack{<ij>, \sigma}} t_{ij,\sigma} n_{i, \bar{\sigma}} c_{i, \sigma}^{\dagger} c_{j, \sigma}\left(1-n_{j, \bar{\sigma}}\right) + h.c.  .\nonumber
\end{align}

$T_{h}$ and $T_{d}$ maintain the number of doubly occupied sites and commute with $H_{U}$. $T_{h}$ transfer one electron from a single-occupied site to an empty one, while $T_{d}$ transfer one electron to a singly occupied one. 
$T_{mix}$ couple the subspaces spanned by $\mathcal{P}$ and $\mathcal{Q}$. We define 
\begin{align}
H=&H_{0}+T_{mix},\nonumber\\
H_{0}=&T_{h}+T_{d}+H_{U}.\nonumber
\end{align}

The effective Hamiltonian $H_{\text{eff}}$ can be obtained by Schrieffer-Wolff transformation to derive in $\mathcal{P}$ by eliminating $\mathcal{Q}$ with $\mathcal{P}H_{\text{eff}} \mathcal{Q}=0$. We seek a canonical transformation $e^{S}$ that decouples $\mathcal{P}$ and $\mathcal{Q}$. Then,  we can derive the effective Hamiltonian by Baker-Campbell-Hausdorff:
\begin{align}
H_{\text{eff}} =& e^{-S} H e^{S} \nonumber\\
=& H +[H,S] + \frac{1}{2} [[H,S],S]+... \nonumber\\
=& H_{0} + \frac{1}{2}[T_{mix}, S] + (T_{mix} + [H_{0}, S]) + \frac{1}{2}[T_{mix}, [H_{0}, S]] + \cdots .\nonumber\
\end{align}
To eliminate the coupling between the subspaces spanned $\mathcal{P}$ and $\mathcal{Q}$, we can choose 
\begin{align}
S=(\mathcal{P} T_{mix} \mathcal{Q} - \mathcal{Q}T_{mix} \mathcal{P})/U. \nonumber\
\end{align}
Then, $T_{mix} + [H_{0}, S]=0$. 

The effective Hamiltonian can be written as 
\begin{align}
H_{\text{eff}} =& H_{0} + \frac{1}{2}[T_{mix}, S],\nonumber\\
         =& T_{h}+T_{d}+H_{U}-\frac{(\mathcal{P} T_{mix} \mathcal{Q})( \mathcal{Q}T_{mix} \mathcal{P})-(\mathcal{Q}T_{mix} \mathcal{P})(\mathcal{P} T_{mix} \mathcal{Q} )}{U},\nonumber\\
         =& H_{0}-\frac{1}{U}(H_{1}+H_{2}).\nonumber
\end{align}
with
\begin{align}
&H_{1}=\sum_{<ij>, \sigma} t_{ij,\sigma} t_{ij,\tau}\left(1-n_{i, \bar{\sigma}}\right) c_{i, \sigma}^{\dagger} c_{j, \sigma} n_{j, \bar{\sigma}} n_{j, \bar{\tau}} c_{j, \tau}^{\dagger} c_{i, \tau}\left(1-n_{i \bar{\tau}}\right), \nonumber\\
&H_{2}=\sum_{<i,j,k>, \sigma, \tau}  t_{i j,\sigma} t_{j k,\tau}\left(1-n_{i, \bar{\sigma}}\right) c_{i, \sigma}^{\dagger} c_{j \sigma} n_{j \bar{\sigma}} n_{j, \bar{\tau}} c_{j \tau}^{\dagger} c_{k, \tau}\left(1-n_{k, \bar{\tau}}\right) .\nonumber
\end{align}
Here, $<i,j,k>$ denotes a summation in which $i \neq k$, and both $i$ and $k$ are nearest neighbor to $j$. For simplicity, we only consider the term of $H_{1}$.

The effective Hamiltonian on the subspace $\mathcal{P}$ is
\begin{align}
H_{\text{eff}}^{P} = \mathcal{P} H_{\text{eff}} \mathcal{P} 
= T_{h} -\frac{1}{U}\mathcal{P} H_{1} \mathcal{P}. \nonumber
\end{align}

We also have 
\[
\mathcal{P} n_{i, \sigma}\left(1-n_{i, \bar{\sigma}}\right) n_{j, \sigma}\left(1-n_{j, \bar{\sigma}}\right) \mathcal{P} =  \mathcal{P} n_{i} n_{j} \mathcal{P}.
\]

The effective Hamiltonian in the subspace spanned by $\mathcal{P}$ can be written as:
\begin{align}
H_{t-J} =  -&\sum_{\substack{\substack{<ij>, \sigma}}} \left( t_{ij,\sigma} \tilde{c}_{i,\sigma}^\dagger \tilde{c}_{j,\sigma} + \text{h.c.} \right) \nonumber\\
+& \sum_{<ij>} \left( J_{\perp}\mathbf{S}^{\perp}_i \cdot \mathbf{S}^{\perp}_{j} +  J_{z} S_{i}^{z} S_{j}^{z}- J_{z}\frac{n_i n_{j}}{4} \right). \nonumber
\end{align}
with $\tilde{c}_{i,\sigma}^\dagger = \left(1-n_{i, \bar{\sigma}}\right) c_{i, \sigma}^{\dagger}$, $\tilde{c}_{i,\sigma} = c_{i, \sigma} \left(1-n_{i, \bar{\sigma}}\right) $ and 
\begin{align}
J_{\perp} =& \frac{4 t_{ij,\sigma}  t_{ij,\bar{\sigma}}}{U} = \frac{4(t^2 - t_A^2)}{U},\nonumber\\
J_{z} =& \frac{2 t_{ij,\uparrow}^{2}+ 2 t_{ij,\downarrow}^{2}}{U} = \frac{4(t^2 + t_A^2)}{U}.\nonumber
\end{align}


In the mean field framework based on the slave boson method, the Hamiltonian can be written as:

\begin{align}
H_{\mathrm{eff}} = &\sum_{\mathbf{k}\sigma} \varepsilon_{\mathbf{k}\sigma} C_{\mathbf{k}\sigma}^\dagger C_{\mathbf{k}\sigma} - J_z \sum_\mathbf{k} \left( C_{\mathbf{k}\uparrow}^\dagger C_{-\mathbf{k}\downarrow}^\dagger \Delta_\eta(\mathbf{k}) + \Delta_\eta^*(\mathbf{k}) C_{-\mathbf{k}\downarrow} C_{\mathbf{k}\uparrow} \right) \nonumber\\
&- 4J_A \sum_\mathbf{k} \left( \Delta_p(\mathbf{k}) C_{\mathbf{k}\uparrow}^\dagger C_{-\mathbf{k}\downarrow}^\dagger  + \Delta_p ^*(\mathbf{k}) C_{-\mathbf{k}\downarrow} C_{\mathbf{k}\uparrow} \right)
+ J_z  |\Delta_\eta|^2 + 8J_A  |\Delta_p|^2.\nonumber
\end{align}
Here, \(\epsilon_{\mathbf{k}}=-2t(\cos k_x+\cos k_y) -\mu\) describes kinetic energy, \(\epsilon_A=-2t_A(\cos k_x - \cos k_y)\) encodes the altermagnetic splitting. 
We consider the extended \(s\)-wave (\(\Delta_s(\mathbf{k})=\Delta_s(\cos k_x+\cos k_y)\)), \(d\)-wave (\(\Delta_d(\mathbf{k})=\Delta_d(\cos k_x-\cos k_y)\))  and \(p\)-wave (\(\Delta_p(\mathbf{k})=\Delta_p(\sin k_x+i\sin k_y)\)) pairing, with the pairing amplitude \(\Delta_{s/p/d}\). 

By introducing the Nambu spinor \(\psi_{\mathbf{k}}=(C_{\mathbf{k}\uparrow},C_{\mathbf{k}\downarrow},C_{-\mathbf{k}\uparrow}^\dagger,C_{-\mathbf{k}\downarrow}^\dagger)^T\), the Hamiltonian can be written as:  

\begin{align}
H_{\mathrm{eff}} =  \frac{1}{2} \sum_{\mathbf{k}} \psi_{\mathbf{k}}^{\dagger} \mathcal{H}(\mathbf{k}) \psi_{\mathbf{k}} 
+ \sum_{\mathbf{k}} \epsilon_{\mathbf{k}} + J_z |\Delta_\eta|^2 + 8J_A |\Delta_p|^2, \nonumber
\end{align}
where the Bogoliubov–de Gennes (BdG) Hamiltonian takes the specific form:
\begin{align}
\mathcal{H} (\mathbf{k}) = \epsilon_k \tau_z + \epsilon_A \sigma_z \tau_z - J_z \Delta_\eta (\mathbf{k}) \sigma_y \tau_y 
+ 8J_A \Delta_p (\sin k_x \sigma_x \tau_x - \sin k_y  \sigma_x \tau_y).\nonumber
\end{align}
At zero temperature, the grand potential takes the form
\begin{align}
\Omega
=\sum_{\mathbf{k}}\epsilon_{\mathbf{k}}
-\tfrac12\sum_{\mathbf{k}}\bigl(E_{\mathbf{k}}^+ + E_{\mathbf{k}}^-\bigr)
+ J_z N |\Delta_\eta|^2 + 8J_A N |\Delta_p|^2. \nonumber
\end{align}
where $E_{\mathbf{k}}^\pm$ denote the quasiparticle excitation spectra of the BdG Hamiltonian. From the underlying $t$–$J$ model, the exchange $J_Z$ grows with $t_A$ and favors spin-singlet channels ($d$-wave or extended $s$-wave), while the anisotropic exchange $J_A$ drives spin-triplet $p$-wave pairing.

\section{Finite‑size scaling analysis}
\label{sec2}
 
 \begin{figure}[h]
\centering
\includegraphics[width=0.9\textwidth]{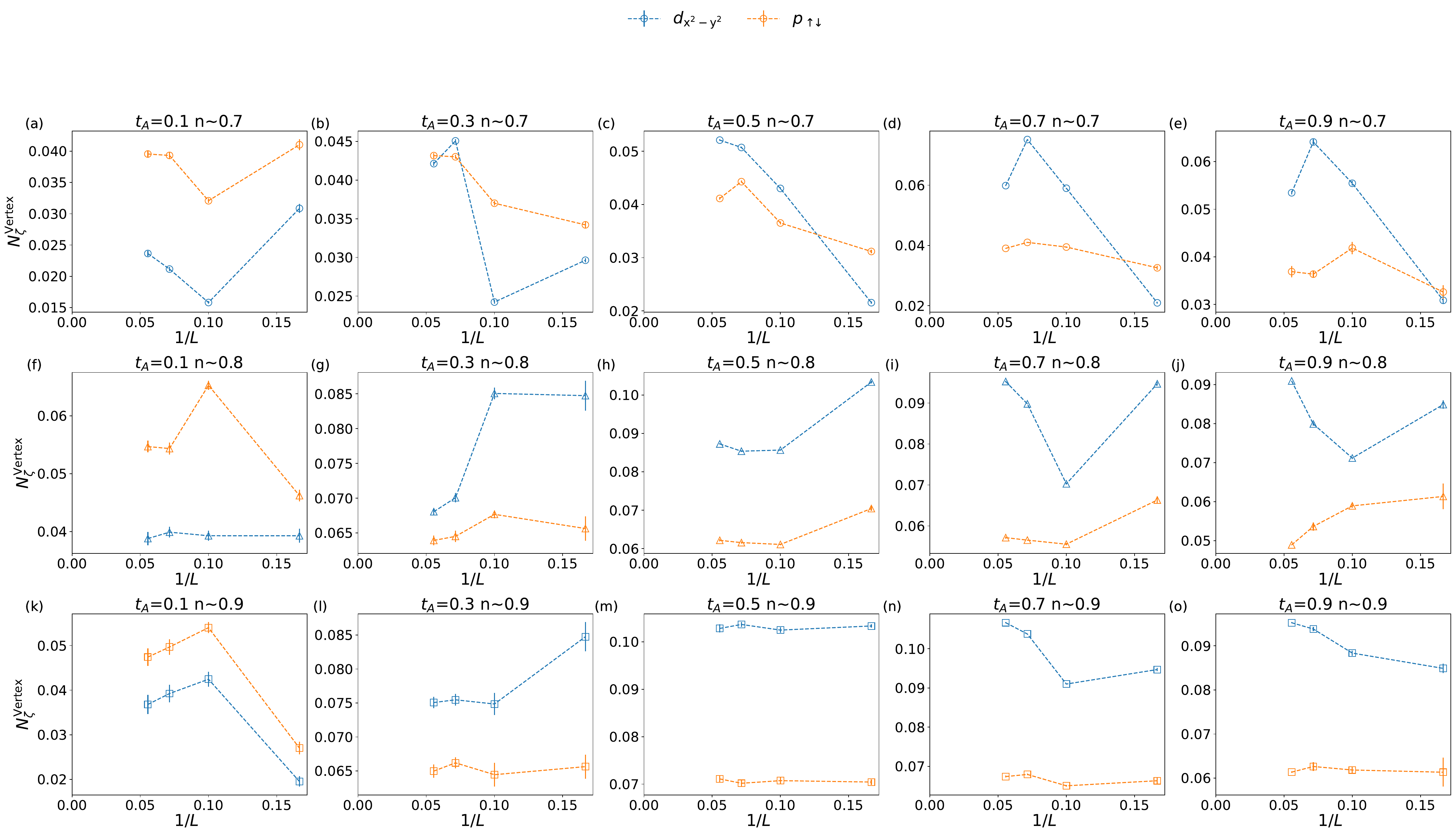}
\caption{For each $t_A$ (columns) and filling $n$ (rows, $n\!\approx\!0.70,\,0.80,\,0.90$), we plot the total vertex pairing
function $N^{\mathrm{Vertex}}_{\zeta}$ versus $1/L$ for $\zeta\in\{d_{x^2-y^2},\,p_{\uparrow\downarrow},\}$ at fixed $U=4$.
Colors encode pairing channels (blue: $d_{x^2-y^2}$, orange: $p_{\uparrow\downarrow}$), and marker shapes encode fillings
(\,$\circ$: $n~0.70$, $\triangle$: $n~0.80$, $\square$: $n~0.90$\,). }\label{supfig1}
\end{figure}

To examine the robustness of our findings, we perform a finite-size scaling analysis of  the vertex pairing functions in different channels. Fig.~\ref{supfig4} present the scaling behavior of the vertex pairing functions 
$N^{\mathrm{Vertex}}_{\zeta}$ with $\zeta \in \{ d_{x^2-y^2}, p_{\uparrow\downarrow}\}$, 
plotted as a function of $1/L$ for representative $t_A$ values at fixed $U=4$. 
The $d_{x^2-y^2}$ channel exhibits a relatively smooth evolution with $1/L$, 
consistent with a robust enhancement of $d$-wave correlations in the thermodynamic limit. 
The $p_{\uparrow\downarrow}$ channel remains weaker but shows a steady finite contribution, 
suggesting that $p$-wave pairing may coexist with $d$-wave correlations in the anisotropic regime. In contrast, the extended-$s$ channel exhibits comparatively weak $1/L$ dependence across the studied parameter space, indicating a background contribution without significant size scaling.

Overall, the finite-size scaling analyses confirm that the enhancement of $d_{x^2-y^2}$ pairing is intrinsic 
and survives in the thermodynamic limit, while  $p_{\uparrow\downarrow}$ channel 
show stronger sensitivity to anisotropy and finite-size effects. These results further support the conclusion that 
altermagnetism favors unconventional pairing tendencies by suppressing long-range antiferromagnetism and stabilizing 
momentum-selective correlations.

\section{Other pairing channel}
\label{sec3}

To complement the discussion of the dominant $d_{x^2-y^2}$ and $p_{\uparrow\downarrow}$ channels, 
we also investigate the behavior of other possible pairing symmetries, 
including the extended-$s$ wave, the spin-triplet $p_{\uparrow\uparrow}$ channel, and the $d_{xy}$-wave channel. 
Fig.~\ref{supfig1} shows the vertex pairing functions $N^{\mathrm{Vertex}}_{\zeta}$ of these channels 
as functions of filling $n$ and hopping anisotropy $t_A$ at fixed interaction strength $U=4.0$.

The extended-$s$ channel exhibits the largest absolute values among the three, 
remaining positive over a broad parameter range. 
Its strength increases at intermediate anisotropy ($t_A \sim 0.5$--$0.7$) 
and fillings $n \gtrsim 0.8$, suggesting that it plays a significant background role, 
though its contribution is not directly driving superconductivity.  

The spin-polarized $p_{\uparrow\uparrow}$ channel stays weakly negative throughout most of the phase space, 
with shallow minima around $t_A \sim 0.3$ and $n \sim 0.8$. 
This indicates that $p_{\uparrow\uparrow}$ correlations remain subleading compared to 
$p_{\uparrow\downarrow}$ or $d_{x^2-y^2}$.  

The $d_{xy}$ channel develops only small negative signals across the $(n,t_A)$ plane, 
with no pronounced enhancement at large anisotropy or near half filling. 
Its contribution is thus minor and size-independent, consistent with being a subleading competitor.  

Taken together, these results establish a clear hierarchy:  
the extended-$s$ channel dominates in magnitude but acts mainly as a competing or suppressive background;  
the $p_{\uparrow\uparrow}$ and $d_{xy}$ channels are both weak and subleading.  
This hierarchy further highlights the robustness of $d_{x^2-y^2}$ pairing as the most favorable channel 
under strong correlations and spin-anisotropic hopping.

\begin{figure}[h]
\centering
\includegraphics[width=0.85\textwidth]{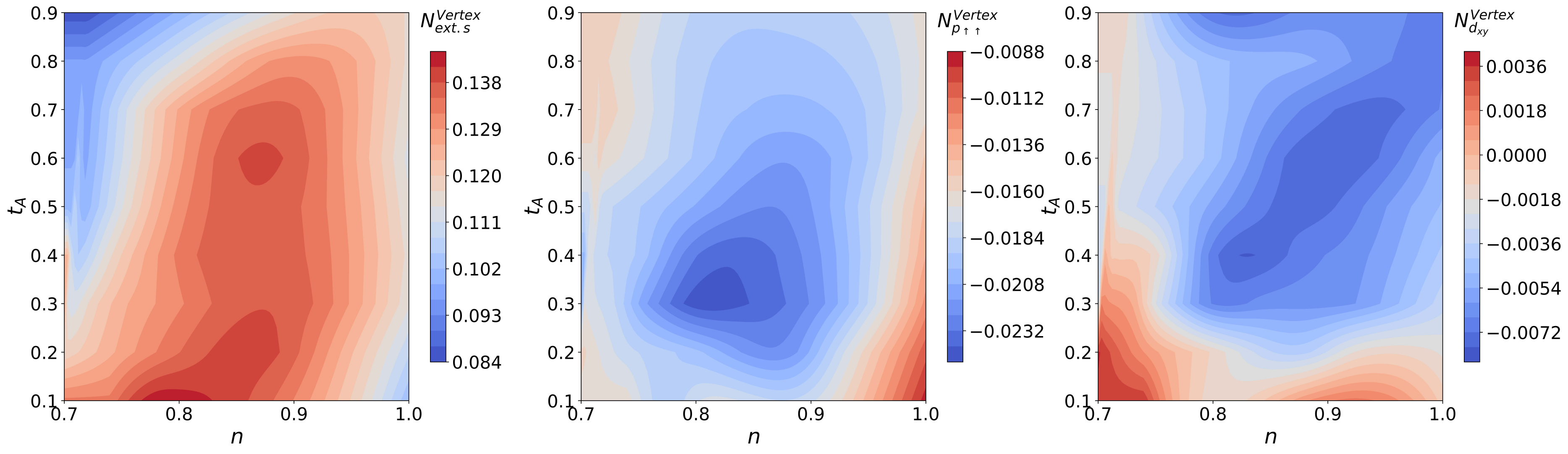}
\caption{
\textbf{Vertex pairing functions $N^{\mathrm{Vertex}}_{\zeta}$ for other channels.}  
Shown are the extended-$s$ (left), spin-triplet $p_{\uparrow\uparrow}$ (middle), 
and $d_{xy}$ (right) channels as functions of filling $n$ and anisotropy $t_A$ at $U=4.0$.  
The extended-$s$ channel reaches the largest magnitude, peaking at intermediate $t_A$ and higher $n$, 
while the $p_{\uparrow\uparrow}$ channel shows weak negative correlations 
and the $d_{xy}$ channel remains small and subleading.}
\label{supfig2}
\end{figure}
\section{Momentum-resolved vertex pairing functions in $d$- and $p$-wave channels}
\label{sec4}
\begin{figure}[h]
\centering
\includegraphics[width=0.65\textwidth]{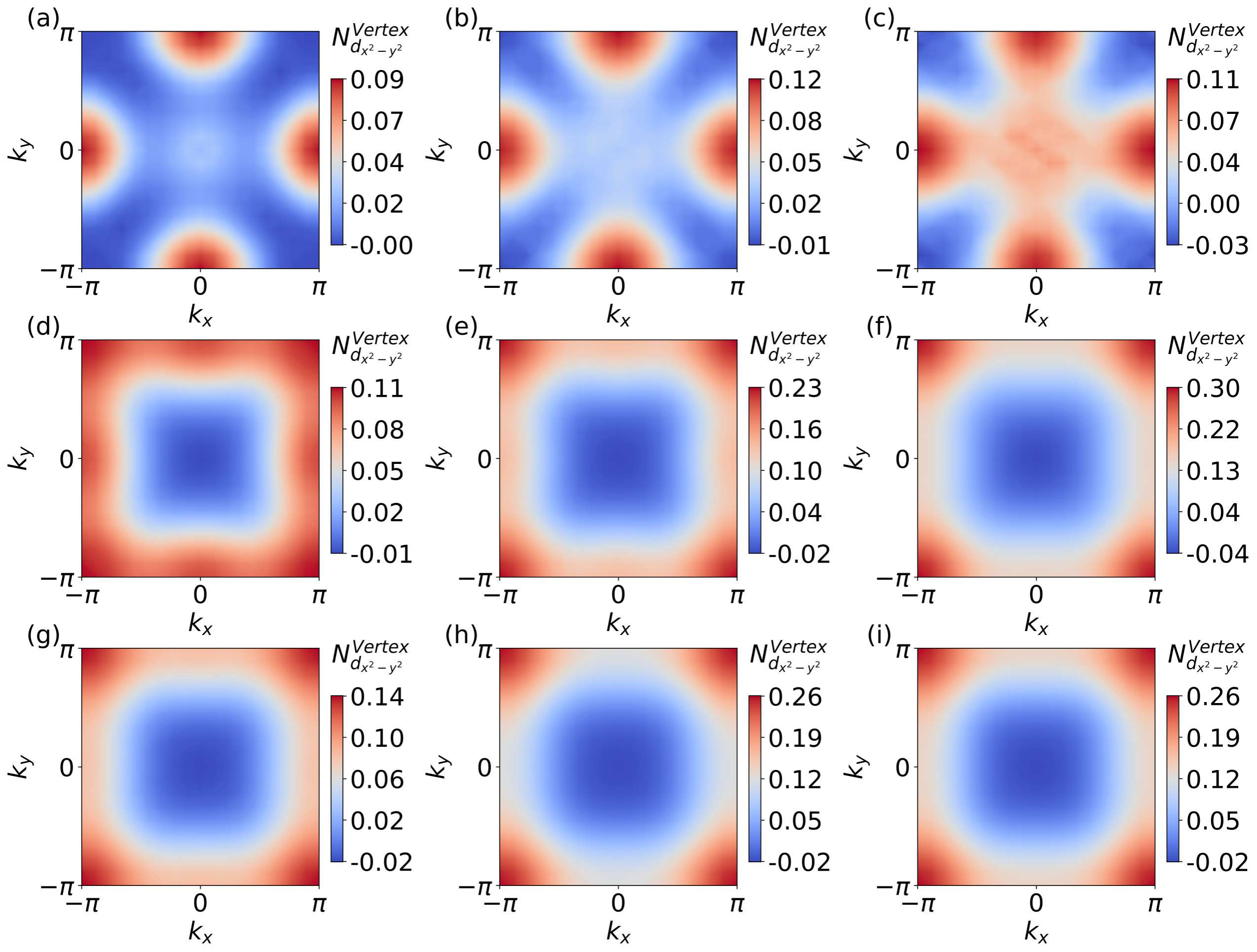}
\caption{\textbf{Momentum-resolved vertex pairing function in the $d_{x^2-y^2}$ channel.}  
Rows correspond to anisotropy $t_A=0.1,0.5,0.9$ (top to bottom), 
and columns to fillings $n \approx 0.7,0.8,0.9$ (left to right).  
The pairing weight evolves from weak and diffuse structures at small $t_A$ 
to strong lobes near $(\pi,0)$ and $(0,\pi)$ at intermediate $t_A$, 
and finally to broad ring-like features at large $t_A$, 
highlighting the robustness of $d$-wave correlations under altermagnetism.}
\label{supfig3}
\end{figure}

\begin{figure}[h]
\centering
\includegraphics[width=0.65\textwidth]{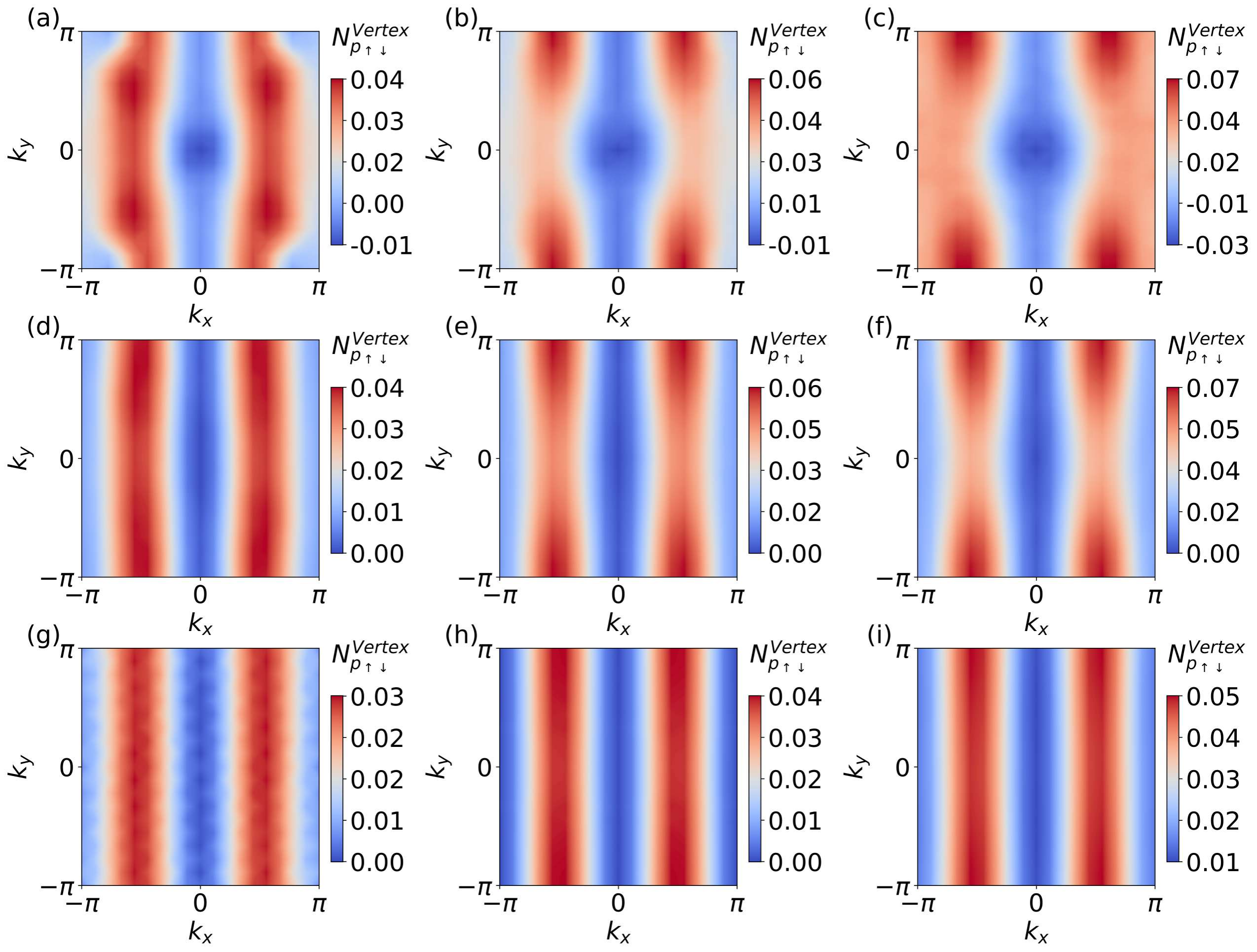}
\caption{\textbf{Momentum-resolved vertex pairing function in the $p_{\uparrow\downarrow}$ channel.}  
Rows correspond to anisotropy $t_A=0.1,0.5,0.9$ (top to bottom), 
and columns to fillings $n \approx 0.7,0.8,0.9$ (left to right).  
The distribution exhibits stripe-like features along $k_x$, 
which sharpen and intensify as $t_A$ increases.  
The filling mainly tunes the magnitude of the contrast, 
with higher $n$ leading to stronger $p$-wave signatures.  
This stripe-dominated structure reflects the quasi-one-dimensional symmetry 
characteristic of the $p$-wave channel.  
}
\label{supfig4}
\end{figure}

 To complement the analysis of spin polarization, 
we present in Fig.~\ref{supfig3} and \ref{supfig4} 
the momentum-resolved vertex pairing functions 
$N^{\mathrm{Vertex}}_{d_{x^2-y^2}}(\mathbf{k})$ 
and $N^{\mathrm{Vertex}}_{p_{\uparrow\downarrow}}(\mathbf{k})$ 
for varying filling $n$ and anisotropy $t_A$. 
As before, the rows correspond to $t_A=0.1,\,0.5,\,0.9$ (top to bottom), 
and the columns to $n \approx 0.7,\,0.8,\,0.9$ (left to right).  

For the $d_{x^2-y^2}$ channel (Fig.~\ref{supfig3}), 
the pairing weight concentrates around the Brillouin-zone boundaries.  
At weak anisotropy ($t_A=0.1$), the structure is relatively diffuse and weak.  
At intermediate anisotropy ($t_A=0.5$), the vertex signal strengthens markedly, 
with enhanced lobes around $(\pi,0)$ and $(0,\pi)$, 
consistent with the amplification of $d$-wave correlations by altermagnetism.  
At strong anisotropy ($t_A=0.9$), the distribution evolves into broad ring-like features, 
suggesting robust $d$-wave character that persists across fillings.  

For the $p_{\uparrow\downarrow}$ channel (Fig.~\ref{supfig4}), 
the distribution is dominated by stripe-like patterns along the $k_x$ direction.  
At low anisotropy, only faint modulations are visible, 
but with increasing $t_A$ the stripes sharpen and extend across the Brillouin zone.  
The filling dependence mainly tunes the intensity of these stripes: 
higher $n$ leads to stronger contrast between the red and blue domains.  
This behavior reflects the quasi-one-dimensional character of the $p$-wave channel, 
whose enhancement is cooperative with $d$-wave pairing near optimal filling.   

\section{ Momentum-resolved SDW and CDW correlation functions}
\label{sec5}
\begin{figure*}[t]
    \centering
    \includegraphics[width=0.95\textwidth]{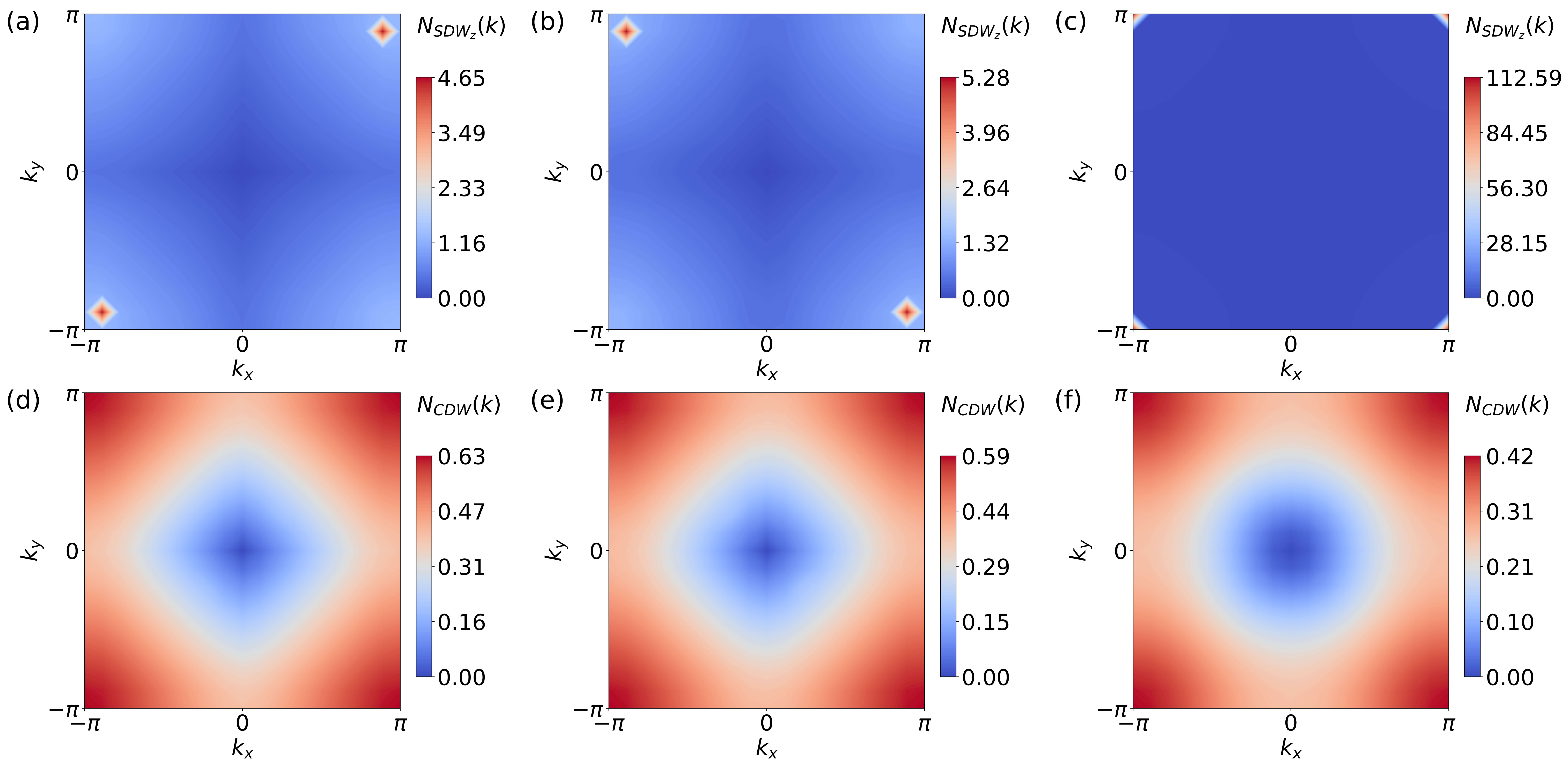}
    \caption{
    \textbf{Momentum-resolved density-wave correlators.}
    (a)--(c) Spin-density structure factor $N_{S_z}(\mathbf{k})$ for three
    representative points along the trajectory in Fig.~2 of the main text,
    showing the suppression of the $(\pi,\pi)$ SDW peak with increasing $t_A$.
    (d)--(f) Charge-density structure factor $N_{\mathrm{CDW}}(\mathbf{k})$
    at the same points, displaying the reduction of boundary-enhanced CDW
    weight as $t_A$ increases.
    }
    \label{supfig6}
\end{figure*}

To complement  $N_{\mathrm{SDW}}$ and
$N_{\mathrm{CDW}}$ shown in the main text, here we provide the spin and charge structure factors $N_{S_z}(\mathbf{k}) = \frac{1}{N}\sum_{i,j}
    e^{i\mathbf{k}\cdot(\mathbf{r}_i-\mathbf{r}_j)}
    \langle S_i^{z} S_j^{z}\rangle, N_{\mathrm{CDW}}(\mathbf{k}) = \frac{1}{N}\sum_{i,j}
    e^{i\mathbf{k}\cdot(\mathbf{r}_i-\mathbf{r}_j)}
    \langle n_i n_j\rangle .$Results for three representative points along the trajectory highlighted in
Fig.~2 of the main text are displayed in Fig.~\ref{supfig6}(a-f).

Fig.~\ref{supfig6}(a-c) show $N_{S_z}(\mathbf{k})$.
At small anisotropy $t_A$, a clear $(\pi,\pi)$ peak is observed, consistent with
strong antiferromagnetic correlations near half filling.
As $t_A$ increases, the peak amplitude is suppressed and the distribution becomes
broader, indicating degraded nesting and weakened SDW tendencies.

Fig.~\ref{supfig6}(d-f) present $N_{\mathrm{CDW}}(\mathbf{k})$ at the same set of points.
At small $t_A$, the spectra show enhanced weight along the Brillouin-zone
boundary, signalling pronounced CDW fluctuations in this doping regime.
Increasing $t_A$ systematically reduces both the boundary weight and the overall
amplitude, demonstrating that spin anisotropy suppresses CDW tendencies.

These momentum-resolved results clarify the origin of the behaviour of the
momentum-averaged quantities $N_{\mathrm{SDW}}$ and $N_{\mathrm{CDW}}$ in the
main text.
In particular, moderate $t_A$ suppresses SDW near $n\!\approx\!1$ and CDW near
$n\!\approx\!0.88$, removing competing density-wave tendencies and allowing the
$d$-wave vertex to reach its maximum.

\section{Momentum-resolved spin polarization as a function of $n$ and $t_A$}
\label{sec6}
\begin{figure}[h]
\centering
\includegraphics[width=0.65\textwidth]{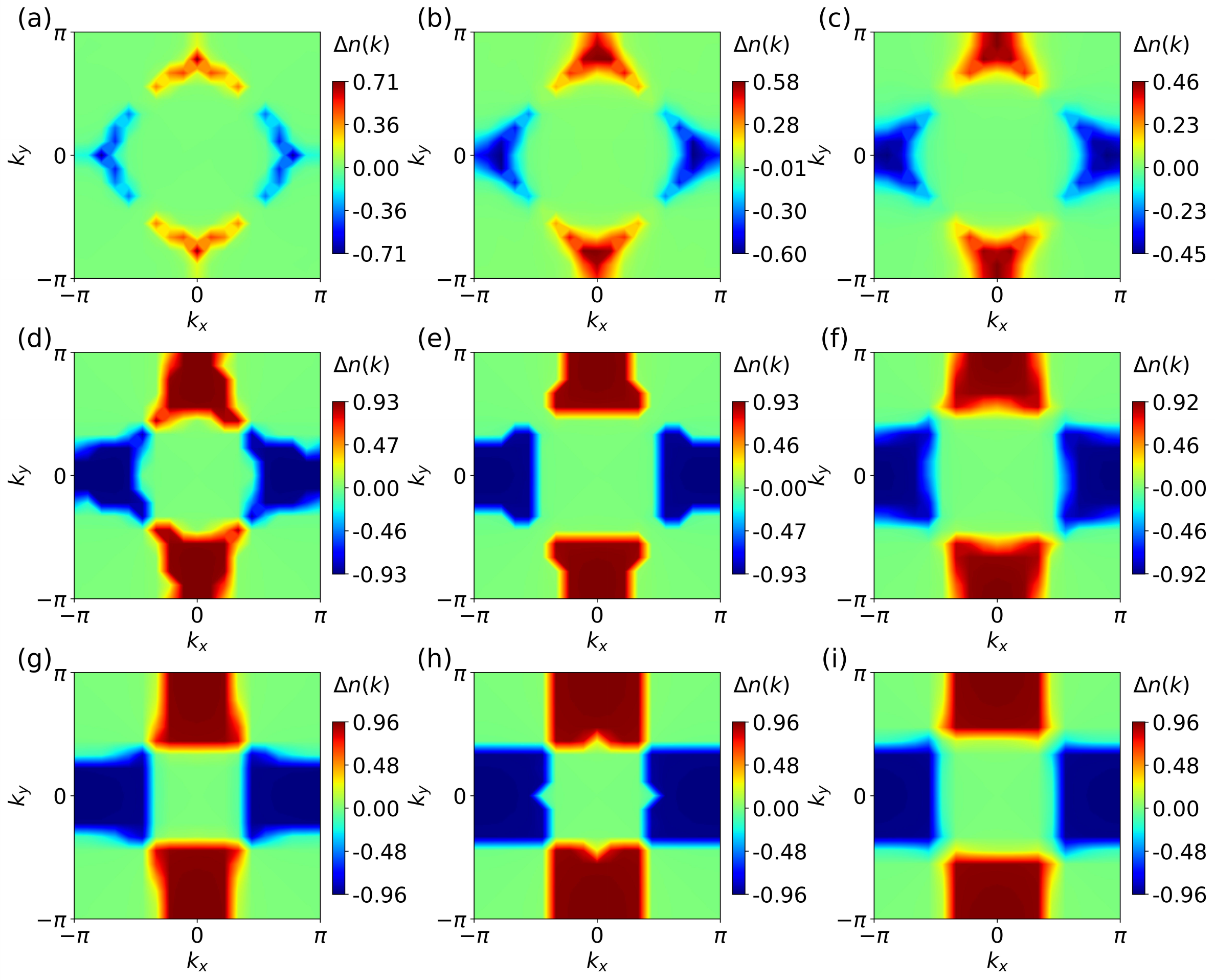}
\caption{\textbf{Momentum-resolved spin polarization $\Delta n(\mathbf{k})$ for different fillings and anisotropies.}  
Rows correspond to anisotropy $t_A=0.1,\,0.5,\,0.9$ (top to bottom), while columns correspond to fillings $n \approx 0.7,\,0.8,\,0.9$ (left to right).  
At weak anisotropy, polarization is confined to localized patches near the Fermi surface;  
at intermediate anisotropy, extended lobes develop along the Brillouin-zone edges;  
and at strong anisotropy, stripe-like domains dominate, reflecting quasi-one-dimensional spin-split bands.  
The enhancement of $\Delta n(\mathbf{k})$ with increasing $n$ and $t_A$ highlights the cooperative stabilization of altermagnetism.  }
\label{supfig5}
\end{figure}

To elucidate the microscopic structure of the altermagnetic phase, 
we compute the momentum-resolved spin polarization $\Delta n(\mathbf{k}) = n_\uparrow(\mathbf{k}) - n_\downarrow(\mathbf{k})$
for representative fillings $n$ and anisotropy parameters $t_A$, as displayed in Fig.~\ref{supfig5}. 
In this figure, the rows correspond to $t_A=0.1,\,0.5,\,0.9$ (from top to bottom), 
while the columns correspond to fillings $n \approx 0.7,\,0.8,\,0.9$ (from left to right).  

For weak anisotropy ($t_A=0.1$), $\Delta n(\mathbf{k})$ remains modest and localized around Fermi-surface hot spots, 
with limited spectral weight and only weak evolution as $n$ increases.  
At intermediate anisotropy ($t_A=0.5$), pronounced polarization lobes emerge and extend along the Brillouin-zone boundaries.  
The imbalance between spin-up and spin-down quasiparticles becomes markedly stronger as $n$ increases, 
signaling the onset of a robust altermagnetic regime.  
At strong anisotropy ($t_A=0.9$), the polarization saturates across large regions of the Brillouin zone, 
yielding stripe-like domains of opposite sign that reflect the quasi-one-dimensional character of the spin-split Fermi surface.  
In this regime, variation in $n$ primarily sharpens the domain boundaries rather than altering the global pattern.  

Overall, the systematic evolution across $n$ and $t_A$ demonstrates their cooperative role:  
while finite $t_A$ is essential for breaking SU(2) symmetry and generating momentum-dependent spin splitting, 
increasing $n$ amplifies the contrast of $\Delta n(\mathbf{k})$, jointly stabilizing the altermagnetic state.  

\section{Robust doped window with $d_{x^2-y^2}$-wave dominance over extended $s$-wave at large $t_A$}
\label{sec7}

To quantify the competition between the leading singlet pairing channels, we examine        the ratio $
R_{d_{x^2-y^2}/s_{\mathrm{ext}}}
\equiv
\frac{N_{d_{x^2-y^2}}}{N_{s_{\mathrm{ext}}}},
$ where $N_{d_{x^2-y^2}}$ and $N_{s_{\mathrm{ext}}}$ denote the integrated pairing vertices
(or integrated effective pairing correlations) in the $d_{x^2-y^2}$ and extended-$s$
channels, respectively.
By construction, $R_{d_{x^2-y^2}/s_{\mathrm{ext}}}>1$ provides a direct criterion that
the $d_{x^2-y^2}$ channel is stronger than the extended-$s$ channel.

\begin{figure}[h]
\centering
\includegraphics[width=0.65\textwidth]{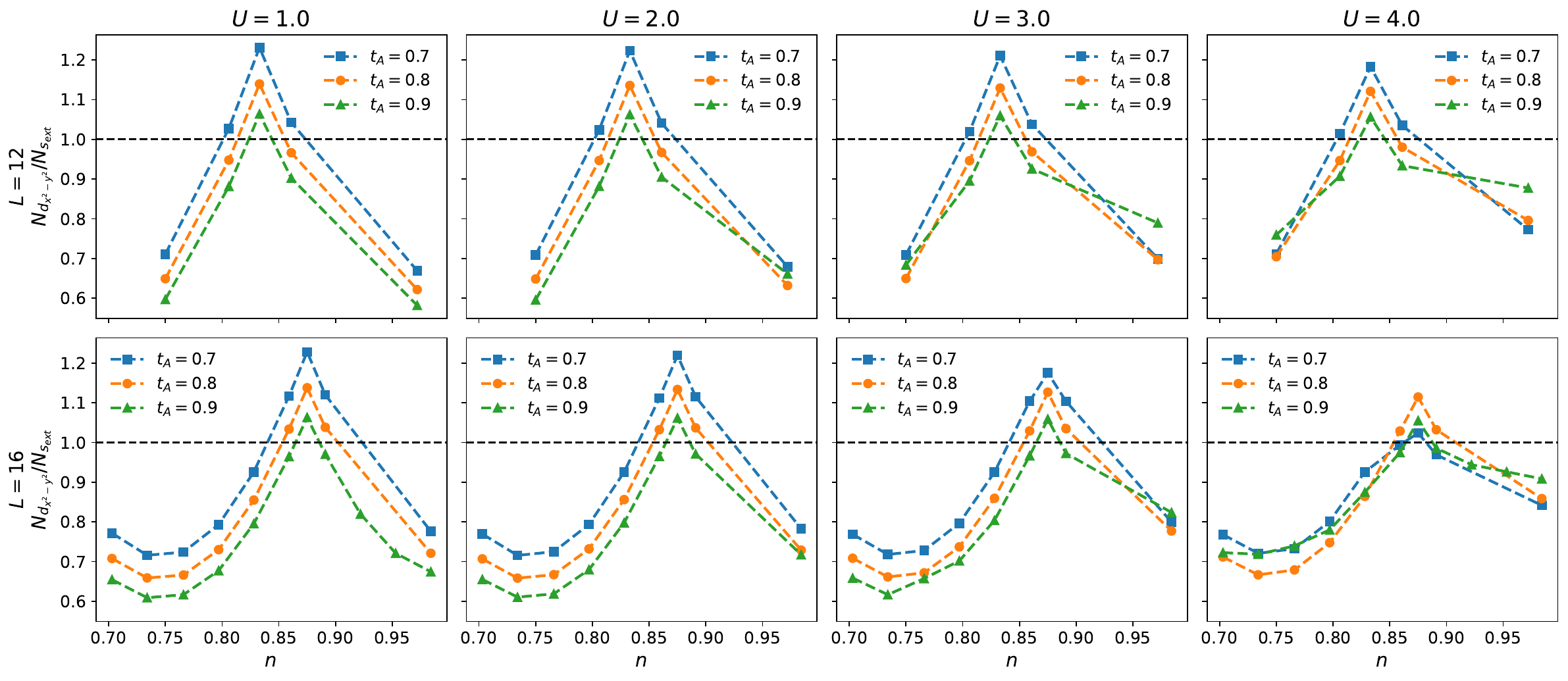}
\caption{Ratio $N_{d_{x^2-y^2}}/N_{s_{\mathrm{ext}}}$ versus filling $n$ for $U=1$--$4$ (columns) and
  $L=12,16$ (rows), in the large-$t_A$ regime $t_A=0.7,0.8,0.9$. The condition
  $N_{d_{x^2- y^2}}/N_{s_{\mathrm{ext}}}>1$ indicates that the $d_{x^2-y^2}$ channel dominates over the
  extended-$s$ channel. For all $U$ and $L$, a finite doped window with
  $N_{d_{x^2-y^2}}/N_{s_{\mathrm{ext}}}>1$ is observed. }
\label{supfig8}
\end{figure}

Fig~\ref{supfig8} shows $R_{d_{x^2-y^2}/s_{\mathrm{ext}}}$ as a function of filling
for interaction strengths $U=1$--$4$, cluster sizes $L=12$ and $16$, and large anisotropic
hopping $t_A=0.7,\,0.8,\,0.9$.
For all values of $U$ and $L$, a finite doped window is observed in which
$R_{d_{x^2-y^2}/s_{\mathrm{ext}}}(n)$ exceeds unity, typically around
$n\simeq 0.83$--$0.90$.
This demonstrates that upon doping, the $d_{x^2-y^2}$ channel consistently overtakes
the extended-$s$ channel in the large-$t_A$ regime.

While the enhancement of $R_{d_{x^2-y^2}/s_{\mathrm{ext}}}$ above unity is moderate
(with peak values of order $1.1$--$1.25$), its appearance is robust against variations
in interaction strength, cluster size, and $t_A$.
Finite-size effects mainly affect the precise peak position and height, but do not alter
the existence of a doped region where the $d_{x^2-y^2}$ channel dominates over
extended-$s$ pairing.

\begin{figure}[h]
\centering
\includegraphics[width=0.65\textwidth]{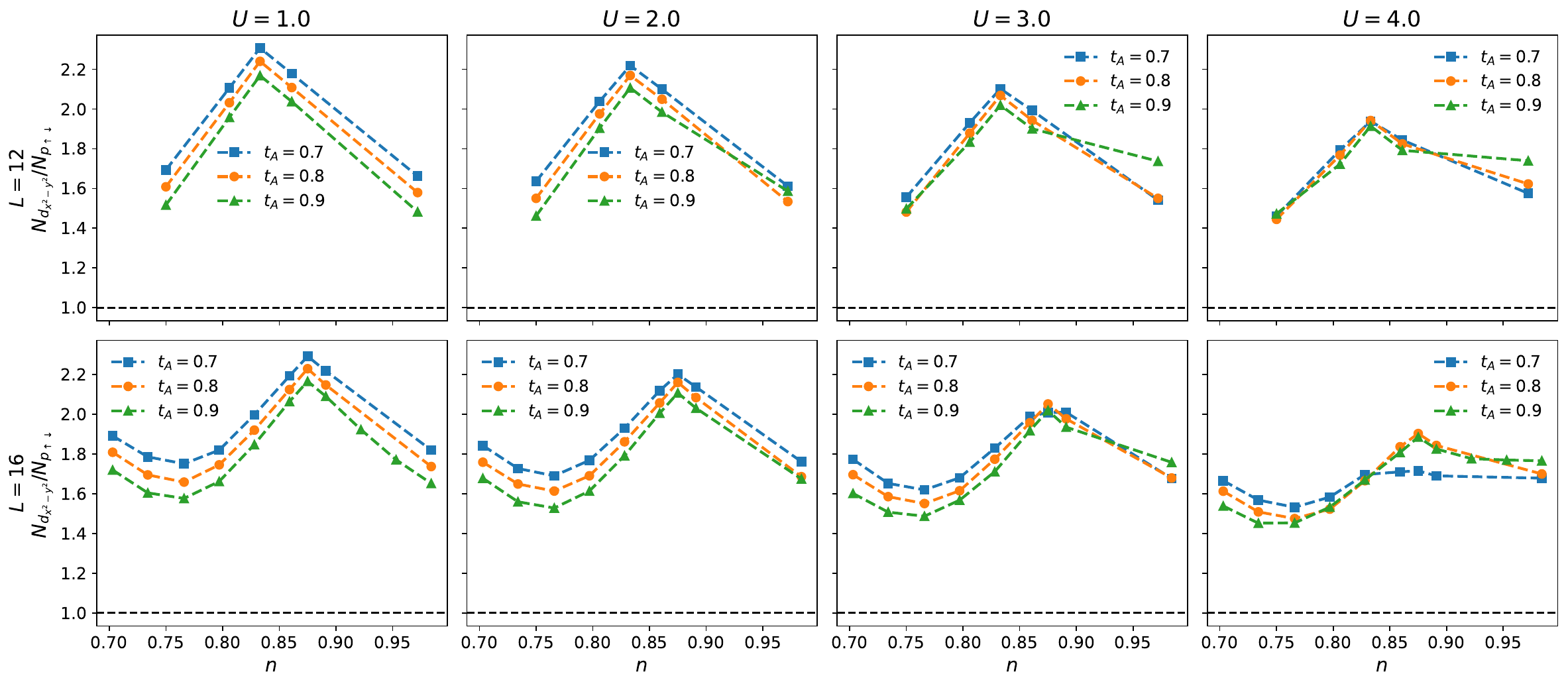}
\caption{Ratio $N_{d_{x^2-y^2}}/N_{p_{\uparrow\downarrow}}$ versus filling $n$ for $U=1$--$4$ (columns) and
  $L=12,16$ (rows), in the large-$t_A$ regime $t_A=0.7,0.8,0.9$. The condition
  $N_{d_{x^2- y^2}}/N_{p_{\uparrow\downarrow}}>1$ indicates that the $d_{x^2-y^2}$ channel dominates over the
  extended-$s$ channel. For all $U$ and $L$, a finite doped window with
  $N_{d_{x^2-y^2}}/N_{s_{\mathrm{ext}}}>1$ is observed. }
\label{supfig9}
\end{figure}

To further establish the dominant nature of the $d_{x^2-y^2}$ channel, we compare it
with the next relevant singlet pairing channel.
Fig~\ref{supfig9} displays the ratio
$N_{d_{x^2-y^2}}/N_{p_{\uparrow\downarrow}}$ over the same parameter range.
In contrast to the marginal enhancement over extended-$s$, the $d_{x^2-y^2}$ channel
is strongly dominant over the competing channel, with
$N_{d_{x^2-y^2}}/N_{p_{\uparrow\downarrow}}$ reaching values between $1.5$ and $2.3$ across a broad doping
window.
This clear separation in magnitude confirms that $d_{x^2-y^2}$ pairing constitutes
the leading superconducting instability in this regime.

Taken together, the combined analysis of
$N_{d_{x^2-y^2}}/N_{s_{\mathrm{ext}}}$ and $N_{d_{x^2-y^2}}/N_{p_{\uparrow\downarrow}}$
demonstrates that, once $t_A$ is sufficiently large, doping generically drives the system
into a robust regime where the $d_{x^2-y^2}$-wave pairing channel becomes dominant.
This dominance is not a finite-size artifact nor restricted to a narrow interaction
range, but persists across all parameters studied here.

\begin{figure}[t]
\centering
\includegraphics[width=0.65\linewidth]{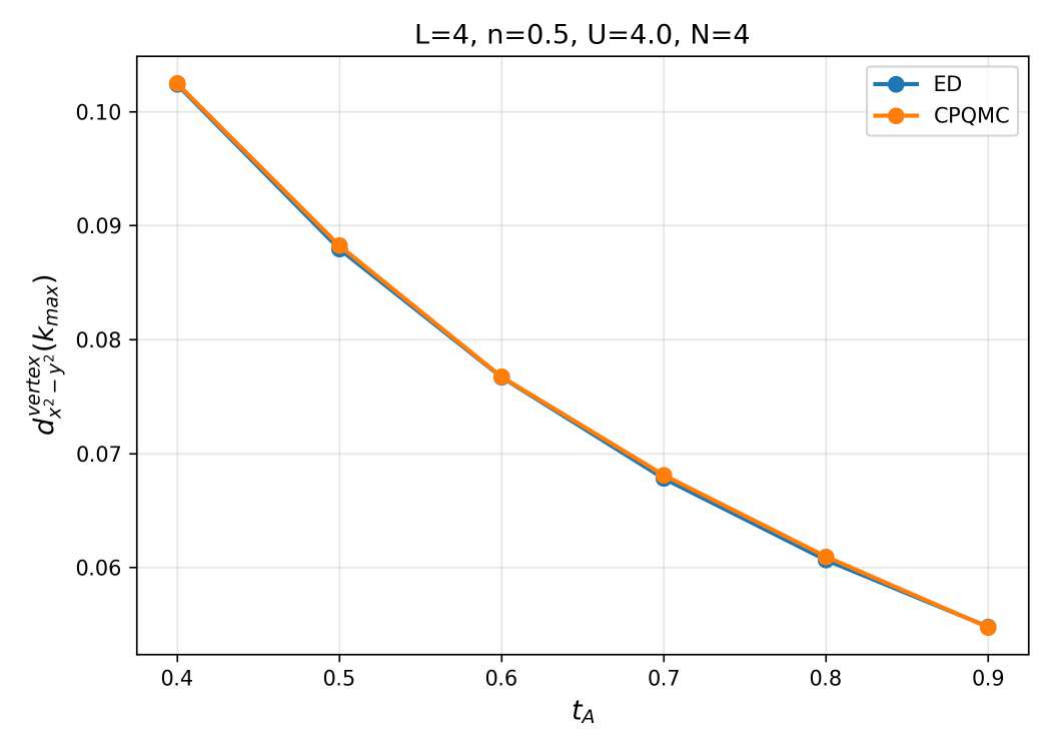}
\caption{
Comparison between exact diagonalization (ED) and constrained-path quantum Monte Carlo (CPQMC)
for the peak value of the \(d_{x^2-y^2}\) vertex pairing correlator,
\(d_{x^2-y^2}^{\mathrm{vertex}}(k_{\max})\), as a function of the altermagnetic anisotropy \(t_A\).
Results are shown for \(L=4\), \(n=0.5\), and \(U=4.0\).
The near-perfect overlap between the two methods demonstrates that CPQMC faithfully reproduces
the ED trend and magnitude in this small-cluster benchmark.
}
\label{supfig10}
\end{figure}
\section{Benchmark of CPQMC against exact diagonalization}
\label{sec8}

As a final validation of our numerical approach, we benchmark the constrained-path quantum Monte Carlo (CPQMC)
calculation against exact diagonalization (ED) on a small cluster where the pairing correlator can be evaluated exactly.
Figure~\ref{supfig10} compares the peak value of the \(d_{x^2-y^2}\) vertex pairing correlator,
\(d_{x^2-y^2}^{\mathrm{vertex}}(k_{\max})\), obtained from ED and CPQMC for \(L=4\), \(n=0.5\), and \(U=4.0\),
as a function of the altermagnetic anisotropy \(t_A\).

The two methods show nearly indistinguishable results throughout the entire range \(0.4 \le t_A \le 0.9\).
In particular, both ED and CPQMC capture the same monotonic decrease of
\(d_{x^2-y^2}^{\mathrm{vertex}}(k_{\max})\) with increasing \(t_A\), with only minute deviations at the level of the plotting symbols.
This agreement indicates that CPQMC accurately reproduces not only the overall magnitude of the pairing response,
but also its detailed anisotropy dependence in the present parameter regime.

Physically, this comparison also shows that, at least in this small-cluster setting,
the \(d_{x^2-y^2}\) pairing vertex is progressively suppressed as \(t_A\) increases.
Since both ED and CPQMC reproduce this trend consistently, the reduction cannot be attributed to a method-dependent bias.
Instead, it reflects an intrinsic change in the pairing environment induced by increasing anisotropy.

Overall, the benchmark in Fig.~\ref{supfig10} provides direct support for the reliability of the CPQMC results
presented in the main text and throughout this Supplemental Material.
\bibliography{sn-bibliography}